%% file: main.tex
\documentclass[aps,pra,reprint,nofootinbib]{revtex4-1}

\usepackage{graphicx}
\graphicspath{{images/}{Images2PDF/}} 
\usepackage{comment}
\usepackage{subfigure}
\usepackage{float}
\usepackage{picture}
\usepackage{xspace}

\usepackage{dcolumn}
\usepackage{bm}
\usepackage[usenames,dvipsnames]{xcolor}
\definecolor{dark-gray}{gray}{0.3}
\usepackage{hyperref}
\hypersetup{
	pdftitle={},
	pdfauthor={Alexander Roth},
	pdfcreator={Alexander Roth},
	pdfproducer={Alexander Roth},
	pdfsubject={},
	pdfkeywords={},
	colorlinks,
	bookmarksopen = true,
	bookmarksnumbered = true,
	citecolor=Blue,
	linkcolor=Blue,
	urlcolor=Blue
 	}

\usepackage{tikz}
\usetikzlibrary{arrows,shapes}
\usetikzlibrary{positioning}
\usetikzlibrary{decorations.pathmorphing}
\usetikzlibrary{decorations.pathreplacing}
\usetikzlibrary{calc}

\tikzset{>=latex}
\usetikzlibrary{shapes.callouts}
\tikzset{
    virtual level/.style = {
        dashed,
        back,
    },
    level/.style = {
        ultra thick,
        black,
    },
    connect/.style = {
        dashed,
        red
    },
    label/.style = {
        text width=2cm
    },
    photon/.style = {
        decorate,decoration={snake,amplitude=.4mm,segment length=2mm },red
    },
    pumping/.style = {
        black!30!green,ultra thick
    },
}

\pgfmathdeclarefunction{gauss}{2}{%
  \pgfmathparse{1/(#2*sqrt(2*pi))*exp(-((\x-#1)^2)/(2*#2^2))}%
}
\pgfmathdeclarefunction{gauss1}{2}{%
  \pgfmathparse{ exp(-((\x-#1)^2)/(2*#2^2))}%
}
\pgfmathdeclarefunction{lorentz}{2}{%
  \pgfmathparse{1/(pi*#2 *(1+((\x-#1)/#2)^2)) }%
}

\usepackage[english]{babel}
\usepackage[utf8]{inputenc}
\usepackage[T1]{fontenc}

\usepackage{amsmath, amscd, amsfonts, amssymb, amsthm} 
\usepackage{ textcomp }
\usepackage{mathtools}
\usepackage{dsfont}
\usepackage[arrow, matrix, curve]{xy}
\usepackage{ifthen}
\usepackage{enumitem}   
\input{Commands}

\pacs{}

\begin{document}

\title{Correlated steady states and Raman lasing in continuously pumped and probed atomic ensembles}

\author{Alexander Roth}
\affiliation{Institute for Theoretical Physics, Institute for Gravitational Physics (Albert Einstein Institute), Leibniz University Hannover, Callinstra{\ss}e 38, 30167 Hannover, Germany}
\author{Kirill S. Tikhonov}%
\affiliation{St. Petersburg State University, 7/9 Universitetskaya nab., St. Petersburg, 199034 Russia}


\author{Klemens Hammerer}%
  \email{klemens.hammerer@itp.uni-hannover.de}
\affiliation{Institute for Theoretical Physics, Institute for Gravitational Physics (Albert Einstein Institute), Leibniz University Hannover, Callinstra{\ss}e 38, 30167 Hannover, Germany}

\date{\today}

\begin{abstract}
\noindent
Spin-polarised atomic ensembles probed by light based on the Faraday interaction are a versatile platform for numerous applications in quantum metrology and quantum information processing. Here we consider an ensemble of Alkali atoms that are continuously optically pumped and probed. Due to the collective scattering of photons at large optical depth, the steady state of atoms does not correspond to an uncorrelated tensor-product state, as is usually assumed. We introduce a self-consistent method to approximate the steady state including the pair correlations, taking into account the multilevel structure of atoms. We find and characterize regimes of Raman lasing, akin to the model of a superradiant laser.  We determine the spectrum of the collectively scattered photons, which also characterises the coherence time of the collective spin excitations on top of the stationary correlated mean-field state, as relevant for applications in metrology and quantum information. 
\end{abstract}

\pacs{Valid PACS appear here}
\maketitle


\section{Introduction}
\label{sec:introduction}

Atomic ensembles coupled to light represent a versatile platform for quantum communication~\cite{Sangouard2011}, for quantum metrology~\cite{pezze_quantum_2018}, optical atomic clocks~\cite{Ludlow2015}, and quantum simulations~\cite{Gross2017}. In particular, the Faraday interaction of light with collective atomic spins has proven to be a powerful tool for realizing a light-matter quantum interface~\cite{hammerer_quantum_2010}, enabling highly efficient quantum control and measurements of atoms~\cite{deutsch_quantum_2010}. It was used in early quantum optics experiments on quantum nondemolition measurements (QND) and has since become a powerful tool for generating spin squeezing of atomic ensembles~\cite{kuzmich_generation_2000,
thomsen_continuous_2002,smith_faraday_2003, smith_continuous_2004,chaudhury_continuous_2006,smith_efficient_2006,Inoue2013,hemmer_squeezing_2021,Takano2009,Sewell2013}. More generally, the Faraday interaction was also utilized to generate and control exotic many-body entanglement~\cite{Behbood2013,Behbood2014}. The exquisite quantum control and long spin coherence time enabled the demonstration of quantum information protocols, including quantum memory and teleportation~\cite{Julsgaard2004,Jensen2010,Sherson2006,Krauter2013} as well as entanglement~\cite{julsgaard_experimental_2001, krauter_entanglement_2011,Thomas2020} and engineered interactions~\cite{Karg2020} between remote macroscopic systems. Further quantum technological applications have been demonstrated in entanglement-enhanced magnetometry~\cite{Wasilewski2010,sewell_magnetic_2012, koschorreck_quantum_2010} and in quantum back-action-evading measurement of motion~\cite{moller_quantum_2017}.

These protocols are usually performed in a \textit{pulsed} mode in the following way: In a first step, a pulse of optical pumping spin-polarizes each atom thus generating an uncorrelated state of atoms with large mean collective polarization. In a second step, a coherent pulse of light reads out a spin projection transverse to the mean polarization through the Faraday interaction. The coherent dynamics achieved in this way can be strong and fast on the scale of relevant decoherence processes, provided the optical depth along the propagation direction of the probe field and the mean polarization of atoms are large~\cite{hammerer_quantum_2010}. In this case, the light-matter interaction exhibits mean-field enhancement, where collective spin excitations are generated on top of the polarised spin state together with collectively scattered photons in specific modes of the light field. The description of this dynamics is based on mean-field theory (MFT), while deviations from the mean-field are described in terms of collective excitations in bosonic spin modes based on e.g. the Holstein-Primakoff approximation \cite{holstein_field_1940, kittel_quantum_1991}. In this way the system is mapped on a number of bosonic modes describing atomic and light degrees of freedom whose time evolution turns out to be linear and integrable~\cite{hammerer_quantum_2010}. Such an account is appropriate for pulsed protocols, where the duration of the probe pulses may be long and even quasi-continuous~\cite{smith_faraday_2003, chaudhury_continuous_2006,  smith_continuous_2004, smith_efficient_2006, kuzmich_generation_2000,
thomsen_continuous_2002,Inoue2013,
hemmer_squeezing_2021} if the probe is sufficiently weak. However, the probe time is ultimately limited by the finite lifetime of atomic spin polarisation, which, aside from other decoherence processes, is set by the probe-induced depumping itself~\cite{hammerer_quantum_2010}.


The regime of \textit{continuous} probing has been considered theoretically as a possibility to generate stationary entanglement among remote atomic ensembles~\cite{Parkins2006,muschik_dissipatively_2011} and in a hybrid system comprising a mechanical oscillator and a collective atomic spin~\cite{Huang2018}. Unconditional steady-state entanglement of atomic ensembles achieved in this way has been reported in~\cite{krauter_entanglement_2011}. In contrast to the pulsed regime outlined previously, optical pumping and probing have to happen at the same time in order to maintain a sufficient stationary atomic polarization supporting a mean field enhancement in the light-atom interaction. In this case, even the steady state defining the mean field  is the result of an interplay between the optical pumping of single atoms and the collective scattering of photons induced by the probe field. As such, it cannot be determined reliably by considering single-atom physics alone. As a result, the model based on the mean field approximation and the Holstein-Primakoff transformation should be adjusted to account for correlations among atoms.



Here, we introduce a self-consistent method based on the cumulant expansion to study continuously pumped and probed atomic ensembles beyond standard mean-field theory. Our model describes a fairly general setup comprised of an ensemble of Alkali atoms with a ground state spin $F$ subject to continuous optical pumping and transverse probing. The probe field is considered to have a linear polarization enclosing an adjustable angle with the axis of optical pumping. This corresponds in particular to the setups studied in~\cite{Julsgaard2004,Jensen2010,Sherson2006,Krauter2013,julsgaard_experimental_2001, krauter_entanglement_2011,Thomas2020}. We give an ab initio derivation of an effective Lindblad master equation for an optically thick ensemble of $N$ spin-$F$ systems, accounting for single atom optical pumping as well as collective scattering events generating correlations among atoms. We solve the master equation for its approximate steady state in a cumulant expansion considering two-particle correlations. By means of the quantum regression theorem, we also determine the spectrum of collectively scattered photons. The width of the corresponding spectral lines determines the coherence time of the spin oscillator associated with collective atomic excitation on top of the correlated mean-field state. We find that the system exhibits features of line narrowing and instabilities associated with transitions to regimes of continuous Raman lasing. These effects depend on the optical depth, which sets the strength of collective scattering relative to individual depumping, but also on the geometry of the setup and in particular the angle among the directions of light and atomic polarization.

We develop our treatment of an optically pumped and probed atomic ensemble by drawing a formal analogy the model of a superradiant laser introduced in~\cite{meiser_prospects_2009,kolobov_role_1993,bohnet_steady-state_2012}. Both systems are described by a Lindblad master equation where single atom dynamics competes with cooperative effects described by collective jump operators. In addition, in both cases, this competition encompasses laser transitions which are well accounted for by an improved mean field theory based on cumulant expansions. However, while the superradiant laser is mostly considered on the basis of a two-level approximation, it is crucial to take into account all Zeeman substates and Clebsch-Gordan weights for Raman transitions in order to cover the physics in a continuously pumped and probed atomic ensemble.

The article is organized as follows: In section \ref{sec:superradiant laser} we first give a brief introduction to the theory of the superradiant laser and then introduce a slightly more general model, which could be considered a superradiant Raman laser. This model exhibits certain features specific to pumped and probed atomic ensemble, but is simple enough for an analytic characterization of its phase regimes. In section \ref{sec:Master equation of continuously pumped and probed Alkali atoms} we derive the master equation for a continuously pumped and probed ensemble of Alkali atoms. We discuss its approximate solution based on a cumulant expansion and explore its features on the basis of the model for the generalized superradiant laser. Finally, in section~\ref{Sec:Conclusion} we summarize and give an outlook for future studies.

\section{Superradiant laser}\label{sec:superradiant laser}

\subsection{Superradiant laser master equation}
\label{sec:Introduction to the superradiant laser}

\begin{figure}[t]
	\centering
	\includegraphics[width=1\columnwidth]{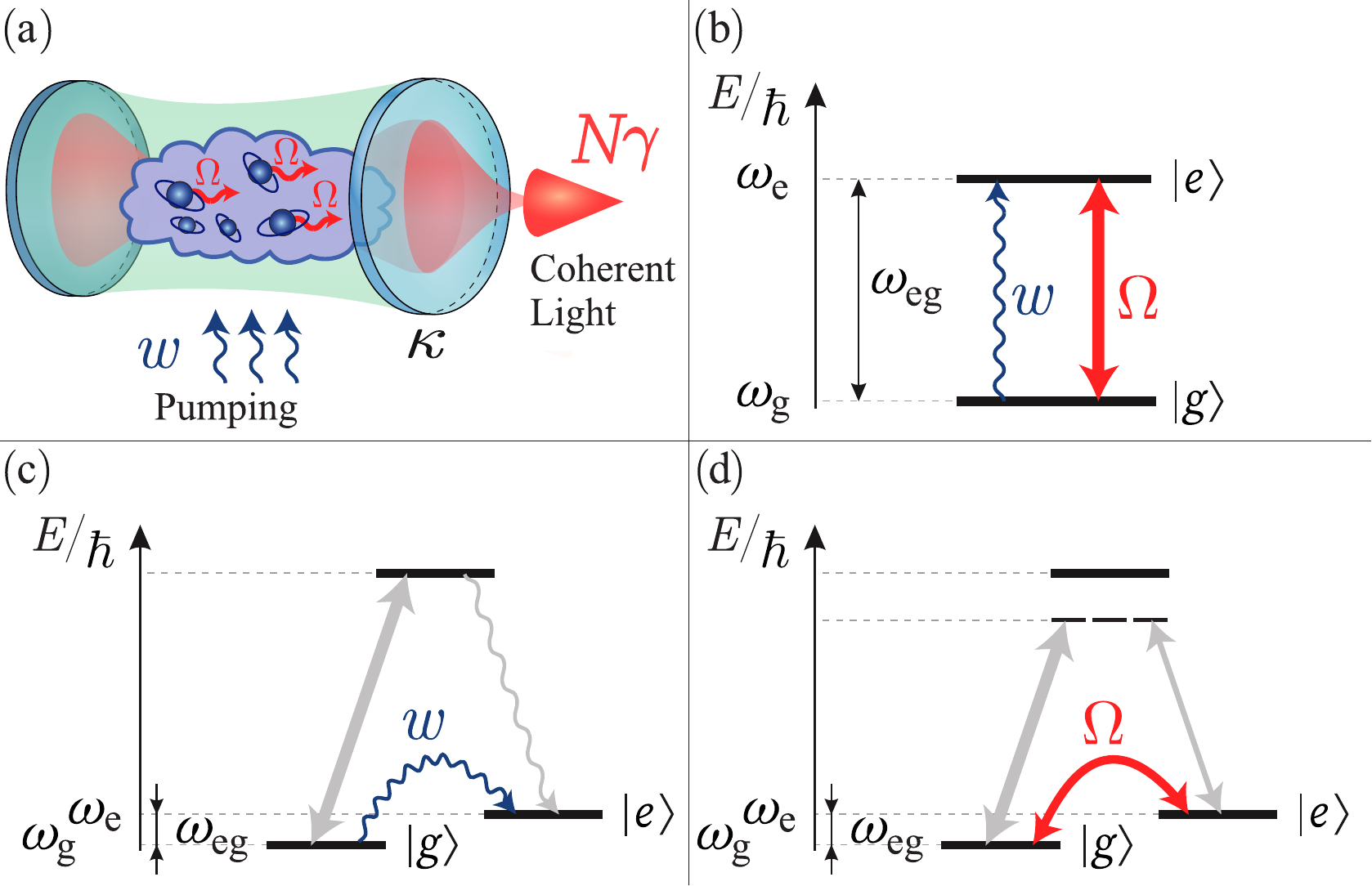}
	\caption{(a) Ensemble of $N$  two-level atoms with single photon Rabi frequency $\Omega/2$ and incoherent pumping   rate $w$, inside a cavity with linewidth $\kappa$. The atoms decays  through the cavity  with rate  $\gamma= \Omega^2/\kappa$. (b) Simplified level scheme of the two-level atoms inside the superradiant laser in (a) with  coherent  atom-cavity interaction  (solid arrows)  and incoherent  pumping rate $w$ (wiggly arrow). (c) Realization of the incoherent pumping process with rate $w$ via a fast-decaying excited state. (d) The coherent  coupling of $\ket g$ and $\ket e$ is achieved through a coherent $\Lambda$-type Raman transition.}
	\label{fig:SRL}
\end{figure}

In this section, we recapitulate the two-level-system model of the superradiant laser and its most general features, which was treated in great detail in~\cite{meiser_prospects_2009,kolobov_role_1993,bohnet_steady-state_2012}. The general setup is shown schematically in figure \ref{fig:SRL}(a). We consider an ensemble of $N$ two-level atoms placed in a cavity with linewidth $\kappa$. The cavity frequency $\omega_{\text{c}}$  is set to be resonant with the atom transition frequency $\atomnu$, i.e. $\omega_{\text{c}}=\atomnu$. The transition between the ground state $\ket g $ and the excited state $\ket e $ couples to the cavity mode $\hat a$ with a single photon Rabi frequency $\Omega/2$, cf. figure~\ref{fig:SRL}(b). The system is described by the Lindblad master equation
\begin{align}
 \frac{d }{dt}\rho &= -i\frac{\atomnu}{2}\C{ J^{\text{z}} +\hat a^\dag \hat a}{ \rho } -i \frac{\Omega}{2}\C{  J^{+}\hat a +  J^{-}\hat a^\dag}{\rho } \breaksign
+ w \sumset[i=1]^N \L{\sigma^{+}_{i}} + \kappa \L{\hat a}
,
\label{02}
\end{align}
 where  $\L{\hat A} = A\rho A^\dag-1/2\C{A^\dag A}{\rho }_+$ is a Lindblad superoperator, $J^{\text{z}}=\sum_{i=1}^{N}\sigma_i^{\text{z}}$ and $J^{\pm}=\sum_{i=1}^{N}\sigma_i^{\pm}$ are collective spin operators, written in terms of the Pauli operator $\sigma^{\text{z}}_i$  and the ladder operators $\sigma^+_i=(\sigma^-_i)^\dag$ for the $i$-th atom.  
 In addition to the cavity decay, the model takes in an incoherent, non-collective pumping process causing population inversion at an effective rate $w$ . This pumping process could correspond e.g. to an additional $\Lambda$-type two-photon process involving a laser assisted excitation followed by a spontaneous emission as shown in figure~\ref{fig:SRL}(c). 
 
 The finite linewidth of the atomic transition could be reflected in an additional Lindblad term in Eq.~\eqref{02}. However, the physics of the superradiant laser relies in particular on the excited state being long lived on the scale of the cavity decay rate. In this limit spontaneous decay plays a minor role and we choose to suppress it here for the sake of clarity. Its role has been dicussed carefully in \cite{meiser_prospects_2009,meiser_steady-state_2010,meiser_intensity_2010} where long lived transitions in Alkaline earth atoms were considered. Another realization of narrow band transitions can be found in $\Lambda$-type Raman transitions as shown in figure~\ref{fig:SRL}(d). This corresponds also to the way the first proof-of-principle realizations of superradiant (Raman) lasing have been achieved~\cite{bohnet_steady-state_2012,bohnet_active_2013}. We note that the following section will expand on this correspondence, and investigate more complicated two-photon transitions and lasing transitions in multilevel atoms.


In contrast to the conventional laser, the superradiant laser relies on collective effects in the atomic medium to store its coherence, instead of relying on the long coherence time of photons inside the cavity \cite{bohnet_steady-state_2012}. Therefore, we consider the atomic ensemble coupling to the light field  in an extreme bad-cavity regime. In this regime the cavity decay is much faster than all other processes, i.e. $\kappa\gg w, \Omega$, and can be adiabatically eliminated \cite{meiser_prospects_2009}, resulting in the permutation invariant master equation, taken here in a frame rotating at the atomic transition frequency $\atomnu$,
\begin{align}
 \frac{d }{dt}\rho& =  w\sumset[i=1]^N \L{\sigma^{+}_{i}} + \gamma \L{ J^{-}}
\label{eq2}
\end{align}
with the rate  $\gamma= \Omega^2/\kappa$  of the collective decay term. 

From \eqref{eq2} the evolution of the expectation values
of $\Mean{\sigma^{\text{z}}_1}$ and $\Mean{\sigma^{+}_1\sigma^{-}_2}$ follows as
\begin{align}
\frac{d}{dt}\Mean{\sigma^{\text{z}}_1} & =w\B{1-\Mean{\sigma^{\text{z}}_1}}
  -\gamma\B{1+\Mean{\sigma^{\text{z}}_1}}
  \breaksign 
  -2(N-1)\gamma\Mean{\sigma^{+}_1\sigma^{-}_2}
,
\notag
 \\
 \frac{d}{dt}\Mean{\sigma^{+}_1\hat\sigma^{-}_2} &=\Bbreakmiddle{ (N-2)\gamma\Mean{\sigma^{\text{z}}_1}-\B{w+\gamma} }\Mean{\sigma^{+}_1\sigma^{-}_2}
 \breaksign +\frac{\gamma}{2}\B{\Mean{\sigma^{\text{z}}_1}+1}\Mean{\sigma^{\text{z}}_1}\label{1},
\end{align}
where we used the cumulant expansion and factorized $ \Mean{\sigma^{+}_1\sigma^{-}_2\sigma^{\text{z}}_3}=\Mean{\sigma^{+}_1\sigma^{-}_2}\Mean{\sigma^{\text{z}}_1}$ and
$\Mean{\sigma^{\text{z}}_1}\Mean{\sigma^{\text{z}}_2}=\Mean{\sigma^{\text{z}}_1}^2$ due to negligible cumulants
$\Mean{\sigma^{\text{z}}_1\sigma^{\text{z}}_2}_c$ and $\Mean{\sigma^{+}_1\sigma^{-}_2\sigma^{\text{z}}_3}_c$, cf~\cite{xu_synchronization_2014}.

\begin{figure}[t]
\centering
\includegraphics[width=0.8\columnwidth]{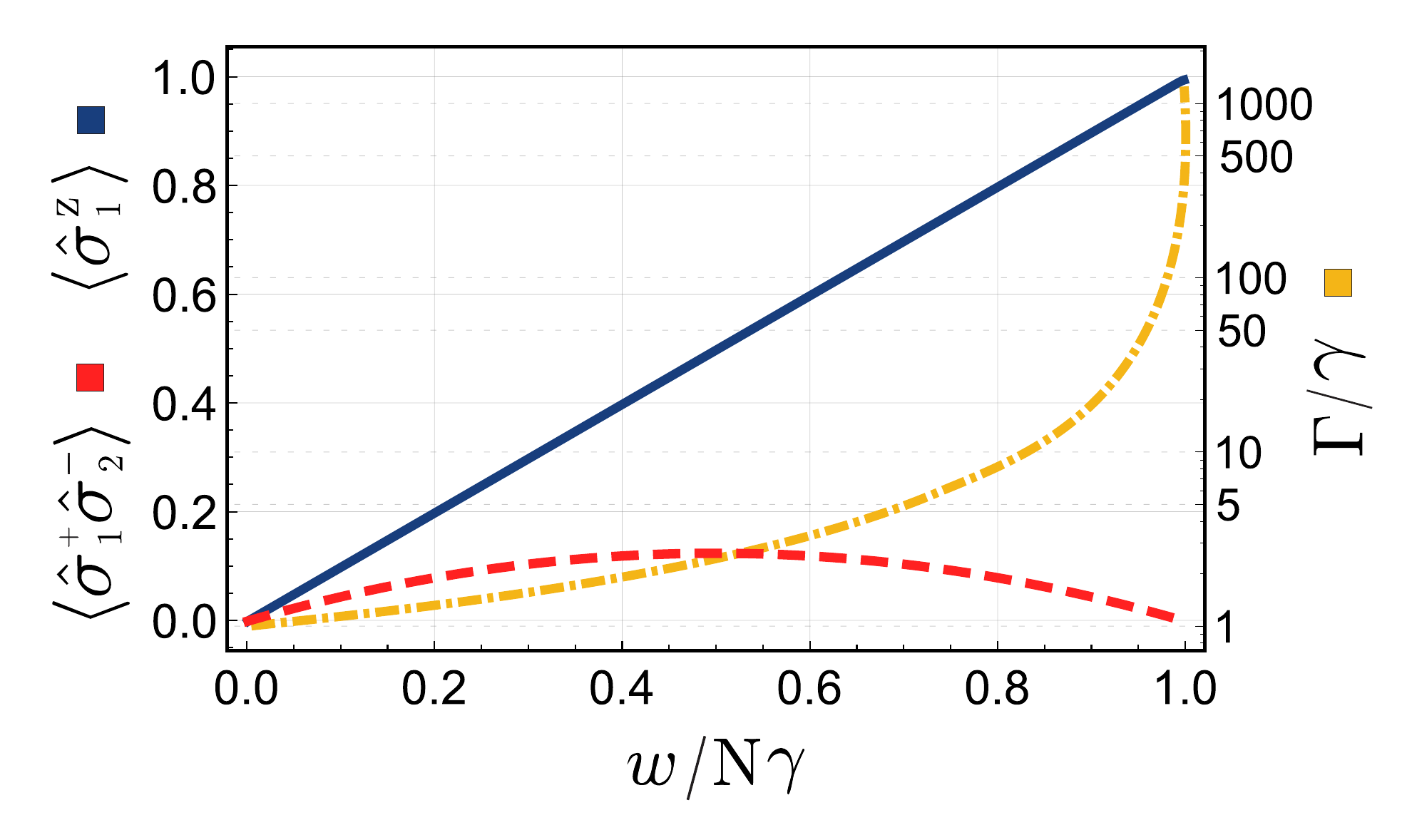}
\caption{The polarization $\Mean{\sigma^{\text{z}}_1}$ (blue solid line), two-atom correlations $\Mean{\sigma^{+}_1\sigma^{-}_2}$ (red dashed line) and the dimensionless linewidth  $\Gamma/\gamma$ (yellow dashed-dotted line) versus dimensionless single-atom pumping rate $w/N\gamma$.
}\label{fig2}
\end{figure}

The steady state expectation values can be obtained by setting the left hand sides of the equations \eqref{1} to zero and solving
the resulting quadratic equation. Figure \ref{fig2} shows the characteristic linearly increasing polarization $\Mean{\sigma^{\text{z}}_1}$ (blue solid line) and the inverted parabola of the correlations $\Mean{\sigma^{+}_1\sigma^{-}_2}$ (red dashed line)  over the single-atom pumping rate $w/N\gamma$. As one can see, the non-zero two-atom correlations for a large atomic ensemble ($N\gg1$) corresponding to the superradiant laser regime exist only when the pumping $w$ fulfills the inequalities
\begin{align}
 \gamma < w < N\gamma.
\label{eq:simple superradiant laser regime}
\end{align}
 At the lower threshold ($w = \gamma$), the pumping overcomes the atomic losses, and the population inversion is established. At the same time, the two-atom correlations build up signifying the onset of superradiance, i.e. atom decay rate $\gamma$ through the cavity is enhanced by a factor proportional to $ N $ (cf. figure \ref{fig:SRL}(a)). It is the minimum condition for lasing, which is in contrast to a conventional laser where the threshold is obtained when the pumping overcomes the cavity losses. At the upper threshold  ($w = N\gamma$), the  two-atom correlations vanish due to the noise imposed by the pumping. Thus, in this case the ensemble consists of random radiators producing thermal light.

The spectrum  of  light leaving the cavity is $S(\omega) = \mathcal{F} [\Mean{ {\hat a}^\dag(t) \hat a(0) }] (\omega)= \frac{\Omega^2}{\kappa^2} \mathcal{F}[ \Mean{ J^+(t) J^-(0)}] (\omega)$
%
%
where $\mathcal{F}$ denotes Fourier transform\footnote{We use the convention $\mathcal{F}[f(t)](\omega)=\frac{1}{\sqrt{2\pi}}\int^{+\infty}_{-\infty}\;dt e^{-i\omega t} f(t)$}.
The equation of motion for the two-time 
collective dipole correlation function
\begin{align}
 \frac{d}{dt} \Mean{ J^+(t) J^-(0)}&= \B{ i \atomnu   - \frac{\Gamma }{2}} \Mean{ J^+(t) J^-(0)}
\end{align}
with $\Gamma=w+\gamma-(N-1)\gamma\Mean{ \sigma^{\text{z}}_1}$ follows from the Quantum Regression Theorem \cite{carmichael_open_1993}.
As a result, the spectrum of the output light of the cavity is Lorentzian with a linewidth $\Gamma$, which is on the order of $\gamma$  \cite{meiser_prospects_2009}. 


At the pumping strength $w_\mathrm{opt}=N\gamma/2$ the atom-atom correlations $\Mean{\sigma^{+}_1\hat\sigma^{-}_2}$ reach their maximum, meaning optimal synchronization of the dipole moment of individual atoms and a corresponding maximal collective atomic dipole moment. This results in the maximal intensity and the relatively narrow linewidth of the output laser light \cite{bohnet_steady-state_2012}. Thus, the superradiant laser regime corresponds to a quite delicate balance between the collective and non-collective processes given in Eq.~\eqref{eq2}.


\subsection{Generalized superradiant laser master equation}
\label{sec:Generalized Superradiant Laser}

\begin{figure}[t]
	\centering
	\includegraphics[width=1\columnwidth]{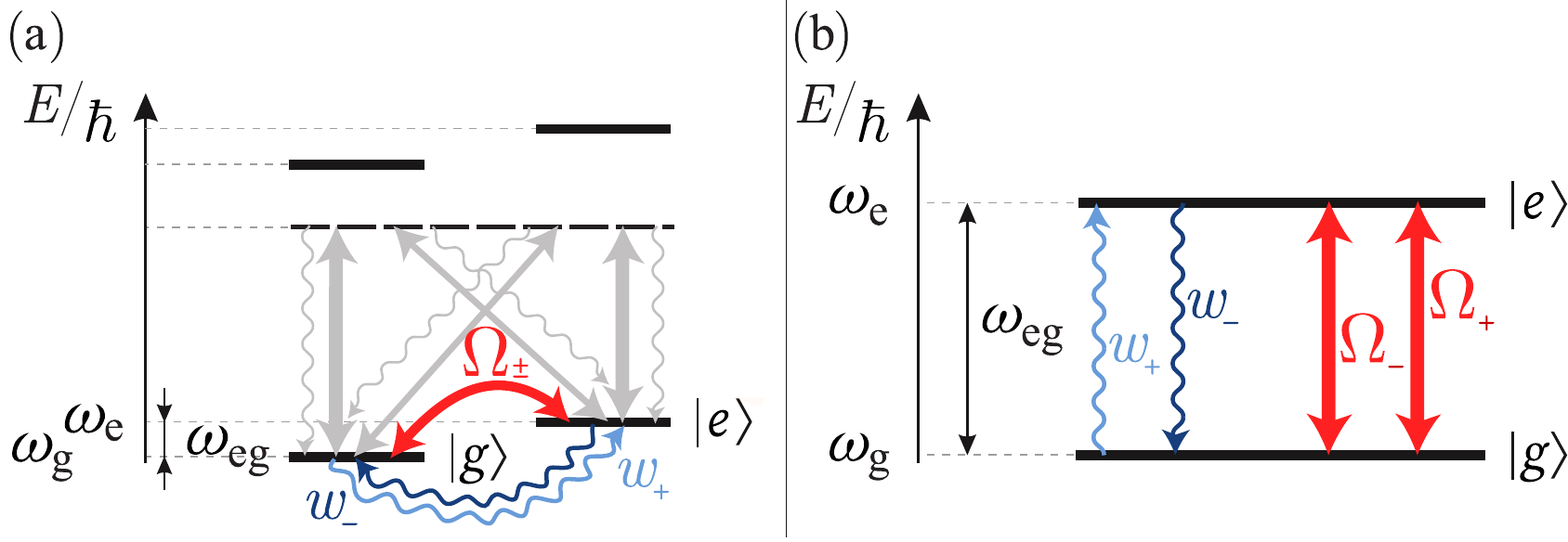}
	\caption{ (a) A level scheme of an atom, which has incoherent  pumping $w_+$ (cf. figure \ref{fig:SRL}(c)), depumping $w_-$, and single photon Rabi frequency  $\Omega_+/2$ (cf.  figure  \ref{fig:SRL}(d)) and counter rotating rate $\Omega_-/2$. (b) This simplified level scheme shows only the relevant levels and processes of (a). It is effectively the level scheme of  figure \ref{fig:SRL}(b) with additional processes for exchanged levels $\ket g \leftrightarrow \ket e$.}
	\label{fig:generalizedSRL}
\end{figure}


We now consider a generalization of the superradiant laser master equation where we allow for additional processes which can arise in more complicated level schemes such as shown in figure~\ref{fig:generalizedSRL}(a). These processes correspond to counter-rotating terms in the picture of the effective two-level system, cf. figure~\ref{fig:generalizedSRL}(b), which may still arise as resonant processes from suitable $\Lambda$-type transitions. Thus, we consider a Jaynes-Cummings like coupling of each atom to the cavity mode at (effective) single photon Rabi rate $\Omega_+$ and an anti-Jaynes-Cummings type interaction at rate $\Omega_-$. Moreover, we also account for individual pumping at rate $w_+$ and at individual depumping from the excited to the ground state at rate $w_-$. All of these processes may arise in double-$\Lambda$ like transitions as shown in figure~\ref{fig:generalizedSRL} and in more complex level schemes as discussed in section~\ref{sec:Master equation of continuously pumped and probed Alkali atoms}.


 After eliminating the excited states, the master equation that accounts for these additional processes in the effective two-level system corresponds to, 
\begin{align}
 \frac{d }{dt}\rho &= -i\frac{\atomnu}{2}\C{ J^{\text{z}}}{ \rho } -i\C{\B{\frac{\Omega_{+}}{2}  J^{+}+\frac{\Omega_{-}}{2}  J^{-}}\hat a + h.c.}{ \rho }
\breaksign + w_+ \sumset[i=1]^N \L{\sigma^{+}_{i}} + w_- \sumset[i=1]^N \L{\sigma^{-}_{i}}+ \kappa \L{\hat a}. \label{eq:genMEQ}
\end{align}
Here, $\atomnu$ accounts for a (possible) energy difference between the two states which physically corresponds to an energy splitting between the two ground states.
Considering the cavity decay as the fastest timescale,  i.e., $\kappa \gg  \Omega_{\pm}, {w_\pm}$,  we perform its
adiabatic elimination as before, resulting in a field that is slaved to the collective atomic dipole of the atomic ensemble, $\hat a \simeq -i (\Omega_{+}J_{+}+\Omega_{-}J_{-})/\kappa$. The master equation for atoms only becomes
\begin{align}
 \frac{d }{dt}\rho &=  w_+ \sumset[i=1]^N \L{\hat \sigma^{+}_{i}} + \gamma_{-} \L{\hat J^{-}} \breaksign +w_-\sumset[i=1]^N \L{\hat \sigma^{-}_{i}} + \gamma_{+} \L{\hat J^{+}}
\label{eq:Generalized superradiant laser}
\end{align}
with rates  $\gamma_{-}= \Omega_{-}^2/\kappa$ and $\gamma_{+}= \Omega_{+}^2/\kappa$ of the collective terms.


The first two terms are identical to the simplified model of the superradiant laser, which were considered in the previous section, while the third and fourth can be regarded as a superradiant laser with interchanged levels. The model considered here thus is unchanged by relabelling $+\leftrightarrow -$. We exploit this symmetry here and assume without loss of generality that the single atom pumping generates population inversion in $\ket{e}$, that is $w_+>w_-$. Furthermore, we are interested in the regime of $w_+,w_-\gg\gamma_-,\gamma_+$ where only collectively enhanced rates $N\gamma_\pm$ are comparable to $w_\pm$. 

We proceed as in the previous section, and derive the evolution of expectation values from \eqref{eq:Generalized superradiant laser}
\begin{align}
\D t  \Mean{\sigma^{\text{z}}_1} &=  w_+\B{1-\Mean{\sigma^{\text{z}}_1}}-w_-\B{1+\Mean{\sigma^{\text{z}}_1}}
\breaksign -2(N-1)\B{\gamma_--\gamma_+}\Mean{\sigma^+_1\sigma^-_2}
\notag
\\
\D t  \Mean{\sigma^+_1\sigma^-_2} &= \Bbreakmiddle{
(N-2)\B{\gamma_--\gamma_+}\Mean{\sigma^{\text{z}}_1}   \breaksign -\B{w_++w_-+\gamma_-+\gamma_+}  } \Mean{\sigma^+_1\sigma^-_2}
\breaksign+\frac{1}{2} \B{ (\gamma_--\gamma_+) +(\gamma_-+\gamma_+)\Mean{\sigma^{\text{z}}_1} ]}  \Mean{\sigma^{\text{z}}_1}
\label{eq:superradiant laser expectation value dynamics}
,
\end{align}
where we factorized $\Mean{\sigma^{\text{z}}_1\sigma^{\text{z}}_2} \approx \Mean{\sigma^{\text{z}}_1}^2 $ and $ \Mean{\sigma^+_1\sigma^-_2 \sigma^{\text{z}}_3}\approx \Mean{\sigma^+_1\sigma^-_2}\Mean{ \sigma^{\text{z}}_1}$ as in \cite{xu_synchronization_2014} assuming negligible cumulants $\Mean[c]{\sigma^{\text{z}}_1\sigma^{\text{z}}_2}   $, $ \Mean[c]{\sigma^+_1\sigma^-_2 \sigma^{\text{z}}_3}$.


\begin{figure}[t]\centering
\includegraphics[width=0.8\columnwidth]{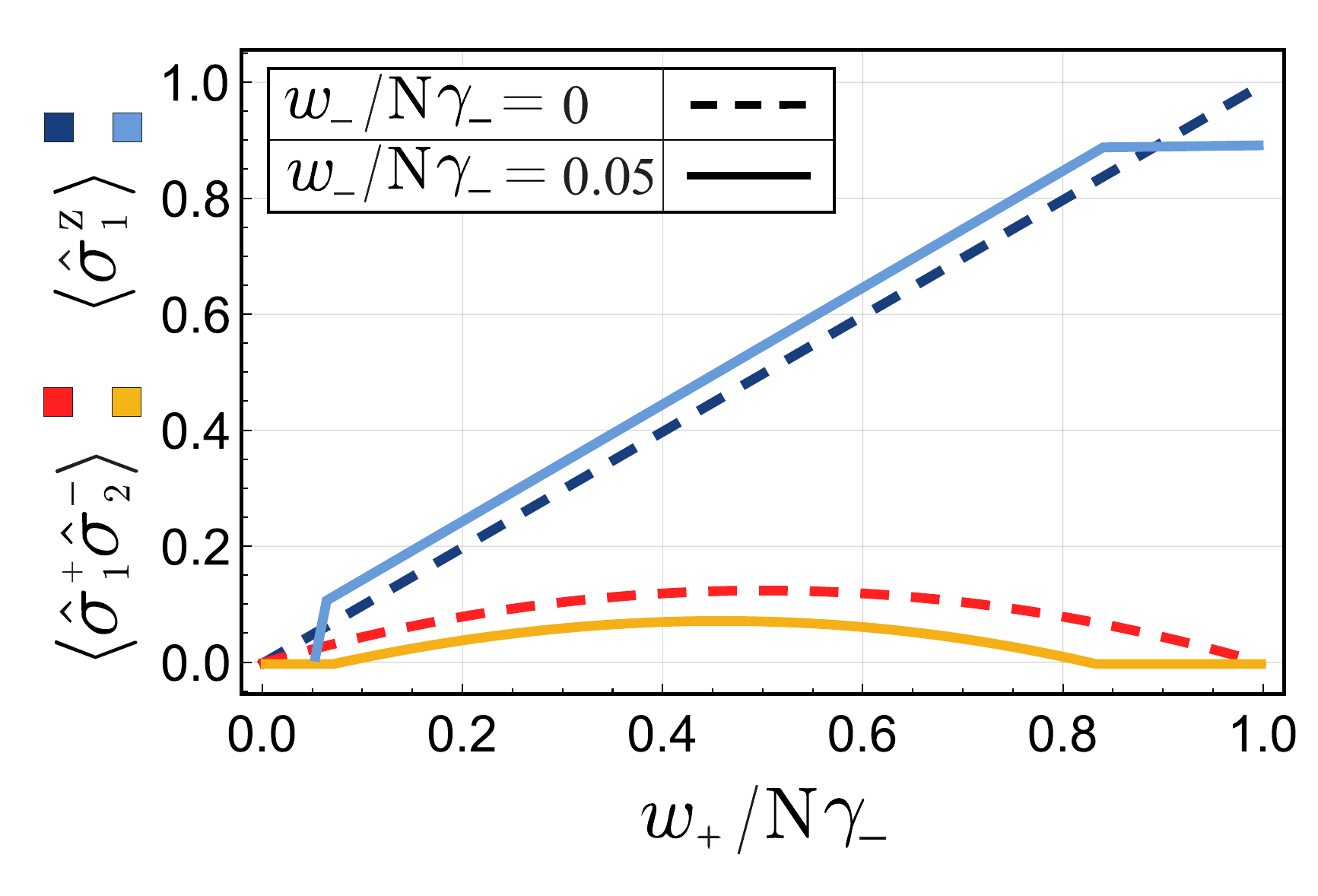} 
\caption{Polarization $\Meansmall{\sigma^{\text{z}}_1}$, and \aa correlation $\Meansmall{\sigma^+_1\sigma^-_2}$ versus single-atom pump rate $w_+$, for vanishing \cuj rate $\gamma_+=0$. 
For comparison, for $w_-/N\gamma_-=0$ (dashed lines) the curves of figure~\ref{fig2} are reproduced.}
\label{fig:superradiant w_+}
\end{figure}

The steady state expectation values can be obtained by setting the left hand sides of Eqs.~\eqref{eq:superradiant laser expectation value dynamics} to zero and solving the resulting quadratic  equation.
The steady state solution of $\Meansmall{\sigma^+_1\sigma^-_2}$ shows that atom-atom correlations, witnessing the regime of superradiant lasing, exist if and only if the \suj rate $w_+$  fulfills the inequality
\begin{align}
 w_+ < N\B{\gamma_- - \gamma_+}  \frac{w_+-w_-}{w_++w_-} - w_-.
\label{eq:Generalized superradiant laser bound}
\end{align}
Here, we assume the limit of a large atom numbers $N\gg 1$ and restrict equations to leading order in $1/N$. Moreover, we assume strong pumping towards level inversion, $w_+\gg \gamma_\pm $. In comparison with \eqref{eq:simple superradiant laser regime}, the threshold condition for the generalized superradiance laser has a nonlinear dependence on $w_+$ for the upper and for the lower bounds. 
We also recall that we took $w_+>w_-$ and conclude that a dominant \cdj rate  $\gamma_- > \gamma_+$ is necessary for superradiance to occur.  

Compared to the model of the superradiant laser, which depended only on the ratio $w/N\gamma$, there are now four independent parameters $N\gamma_\pm$ und $w_\pm$. It will be useful to discuss the steady state physics in terms of the ratios of single atom pumping to collective decay, $w_+/N\gamma_-$, collective excitation to collective decay, $\gamma_+/\gamma_-$, and single atom depumping to collective decay, $w_-/N\gamma_-$.

\begin{figure}[t]\centering
\includegraphics[width=\linewidth]{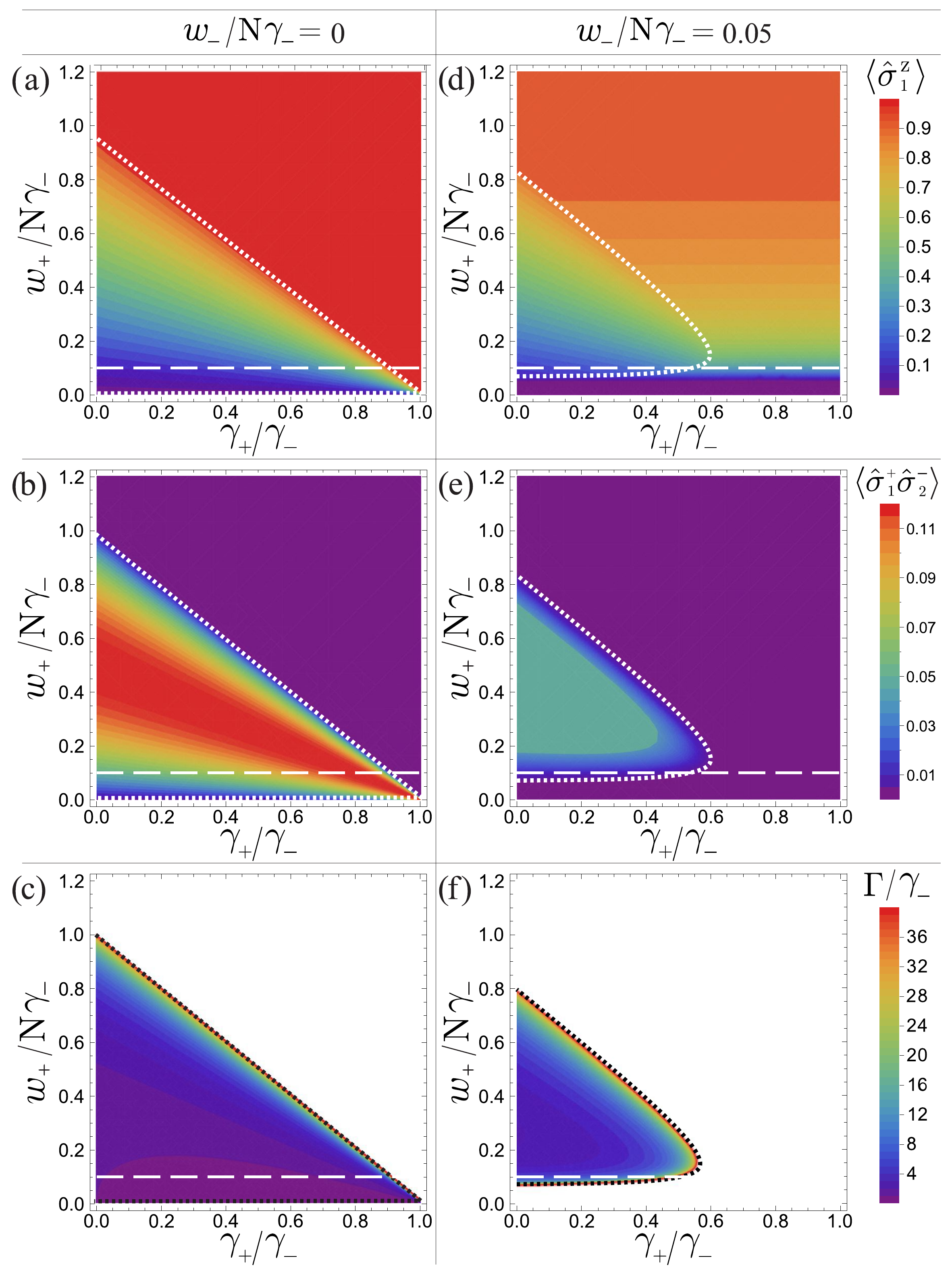}
\caption{Polarization $\Meansmall{\sigma^{\text{z}}_1}$, atom-atom correlation $\Meansmall{\sigma^+_1\sigma^-_2}$, and full-width at half maximum of the Lorentz peak $\Gamma$ (top to bottom row) versus collective excitation rate $\gamma_+/\gamma_-$ and single-atom pump rate $w_+/N\gamma_-$. The  \sdj rate  is  $w_-=0$ in (a), (b), (c)   and  $w_-/N\gamma_-=0.05 $ in (d),  (e), (f). The   dashed lines at  $w_+/N \gamma_-=0.1$ correspond to the parameters in  in figure \ref{fig:superradiant gamma_+}. The  dotted lines are given by \eqref{eq:Generalized superradiant laser bound}, giving an  envelope of  the superradiant lasing regime in leading order in $1/N$.}
\label{fig:superradiant contours}
\end{figure}

\begin{figure}[t]\centering
\includegraphics[width=0.8\columnwidth]{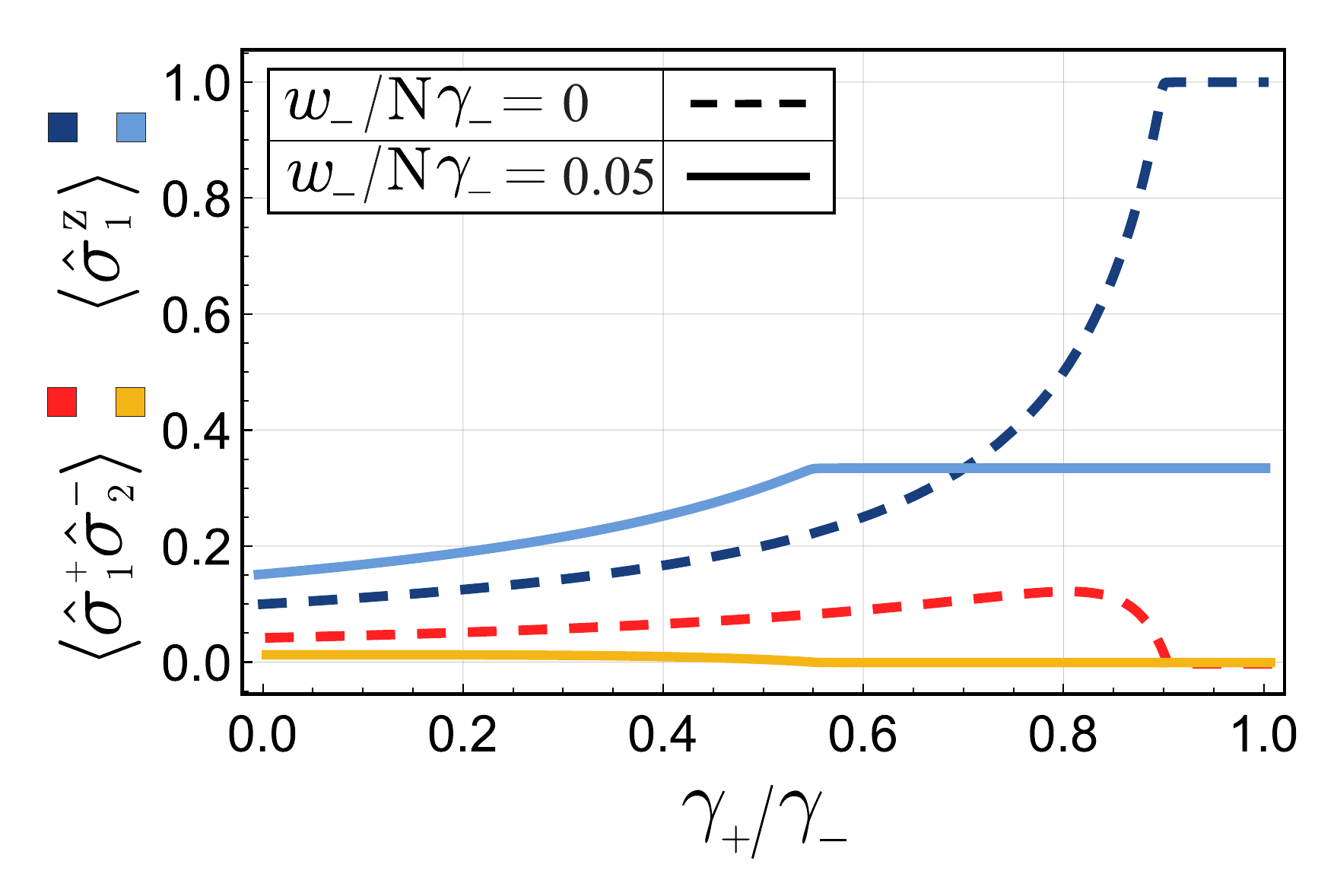}
\caption{Polarization $\Meansmall{\sigma^{\text{z}}_1}$, and \aa correlation $\Meansmall{\sigma^+_1\sigma^-_2}$ versus \cuj rate $\gamma_+/\gamma_-$, corresponding to the dashed lines in  figure~\ref{fig:superradiant contours}. Varying $\gamma_+$ scans through the superradiant regime. For  $\gamma_+$ close  below the upper threshold of the superradiant regime  \eqref{eq:Generalized superradiant laser bound}, the  polarization and \aa correlation are strongly dependent on   $\gamma_+$, and therefore very sensitive to small changes. 
$w_+/N\gamma_-=0.1$ }
\label{fig:superradiant gamma_+}
\end{figure}
Figure~\ref{fig:superradiant w_+} illustrates the case where $\gamma_+/\gamma_-=0$, and shows atomic polarization and dipole correlations versus $w_+/N\gamma_-$, in analogy to what was shown for the superradiant laser in figure~\ref{fig2}. The overall behaviour is similar, but in comparison, the superradiant regime is somewhat reduced for nonzero single atom depumping $w_-/N\gamma_-$ (solid line), as is to be expected.
Figure~\ref{fig:superradiant contours} provides a more complete overview, and shows the steady state polarization  $\Meansmall{\sigma^{\text{z}}_1}$ and atom-atom correlation $\Meansmall{\sigma^+_1\sigma^-_2}$ versus  collective excitation rate $\gamma_+/\gamma_-$ and individual pumping $w_+/N\gamma_-$. The left (right) column of figure~\ref{fig:superradiant contours} refers to vanishing (nonzero) single atom depumping $w_-/N\gamma_-$. The figure illustrates the superradiant domain and shows that it is excellently characterized by condition~\eqref{eq:Generalized superradiant laser bound}. Most importantly, figure~\ref{fig:superradiant contours} reveals a rich dependence of the steady state properties on the ratio of collective excitation and decay ratios $\gamma_+/\gamma_-$. Corresponding cuts along this axis are shown in figure~\eqref{fig:superradiant gamma_+}.
%
%
The behaviour of the system along these will be of importance for our discussion of multilevel atoms in the next section, where we will show that geometrical aspects of the light-matter interactions determine the ratio of the rates $\Omega_\pm$ and with it the ratio of $\gamma_\pm$.
The   maximal  \aa correlations are
\begin{align*}
\max_{w_+} \Mean{\sigma^+_1\sigma^-_2} 
= \frac 1 8 - \frac{w_- }{N \gamma_+  }   \frac{1 }{  \frac{\gamma_-}{\gamma_+} -  1}   
\end{align*}
in leading order in $1/N$, for $\gamma_+\geq \gamma_-$ at the optimal pumping strength $w_{+,\text{opt}}= \fracsmall{N  \B{\gamma_- - {\gamma_+} }}{2 }-w_-$. 


In the generalized superradiant laser the linewidth is still on the order of the atomic linewidth $\gamma_-$, even though the ensemble is incoherently pumped with a much stronger rate $w_+$. We will now show that we essentially have  two superradiant lasing  transitions, $\ket e \to \ket g$ and $\ket g \to \ket e$,  radiating at the same time  with identical linewidth of the order of $\gamma_-$, but different intensities.

As before, the spectrum of the output light $S(\omega) = \mathcal{F} [\Mean{ {\hat a}^\dag(t) \hat a(0) }] (\omega)= \frac{\gamma_+}{\kappa} \mathcal{F}[ \Mean{ J^+(t) J^-(0)}] (\omega)+\frac{\gamma_-}{\kappa} \mathcal{F}[ \Mean{ J^-(t) J^+(0)}] (\omega)$ is evaluated using the \QRT \cite{gardiner_quantum_2004} based on the equation of motion
\begin{align*}
\D t \Mean{J^-(t) J^+(0)}&= \B{ -\i \atomnu   - \frac{\Gamma }{2}}   \Mean{J^-(t) J^+(0)}.
\end{align*}
with linewidth $\Gamma=\gamma_- + \gamma_+  + w_+ + w_- - (N-1)\B{\gamma_- - \gamma_+} \Mean{\sigma^{\text{z}}_1}$.
The corresponding spectrum is given by two  Lorentz functions at $\pm \atomnu$ with heights $S_\pm$ and identical linewidth (full-width at half maximum)  $\Gamma$. 
The linewidth $\Gamma $ for $w_+>w_-$ to leading order $1/N$ is
\begin{align*}
\frac{ \Gamma }{\gamma_-} &\approx 
   \frac{W_+ + W_+W_- (W_- - W_+W_- -1)   }{  \B{W_+-1}\B{W_--1}W_-  } \B{1-\frac{\gamma_+}{\gamma_-}  }
   ,
\end{align*}
where we defined  the dimensionless variables
\begin{align*}
W_\pm :=   \frac{\B{w_+\pm w_-}\B{w_++w_-}}{ N\B{\gamma_- - \gamma_+}  \B{w_+-w_-}  }.
\end{align*}
The linewidth $\Gamma$  is shown in figures~\ref{fig:superradiant contours}, panel (c) and (f). We see that the generalized superradiant laser preserves the  remarkable feature of the superradiant laser -- the linewidth on the order of  the \eadr $\gamma_-$ -- even for  non-vanishing $\gamma_+$. And even with  additional \sdj rate  $w_-$  the linewidth increases only slightly (see figure~\ref{fig:superradiant contours}(f)).  The spectrum exhibits two asymmetric peaks at the sideband frequencies $\omega=\pm\gennu$. The ratio of the sideband intensities $S_\pm$ is given by
\begin{align}
\frac{S_+}{S_-} &= \frac{\gamma_+}{\gamma_-} \frac{1}{1-\frac{\Mean{J^{\text{z}}}}{\Mean{J^+J^-}}}
\approx  \frac{\gamma_+}{\gamma_-}
\label{eq:general superradiant maxima ratio}
\end{align}
for $N\gg 1$ in the superradiant regime, which is simply the ratio of collective emission rates at the sidebands. We want to point out that equal Lorentz peak height is not possible, because the superradiant condition \eqref{eq:Generalized superradiant laser bound} can not be fulfilled for $\gamma_+=\gamma_-$.

The model of the generalized superradiant laser will be a helpful reference to understand the physics of continuously pumped and probed atomic ensembles, which will be treated in the following.

\section{Continuously pumped and probed atomic ensembles}
\label{sec:Master equation of continuously pumped and probed Alkali atoms}

\begin{figure}[t]\centering
	\includegraphics[width=1.0\linewidth, trim= 0 .7cm 0 0]{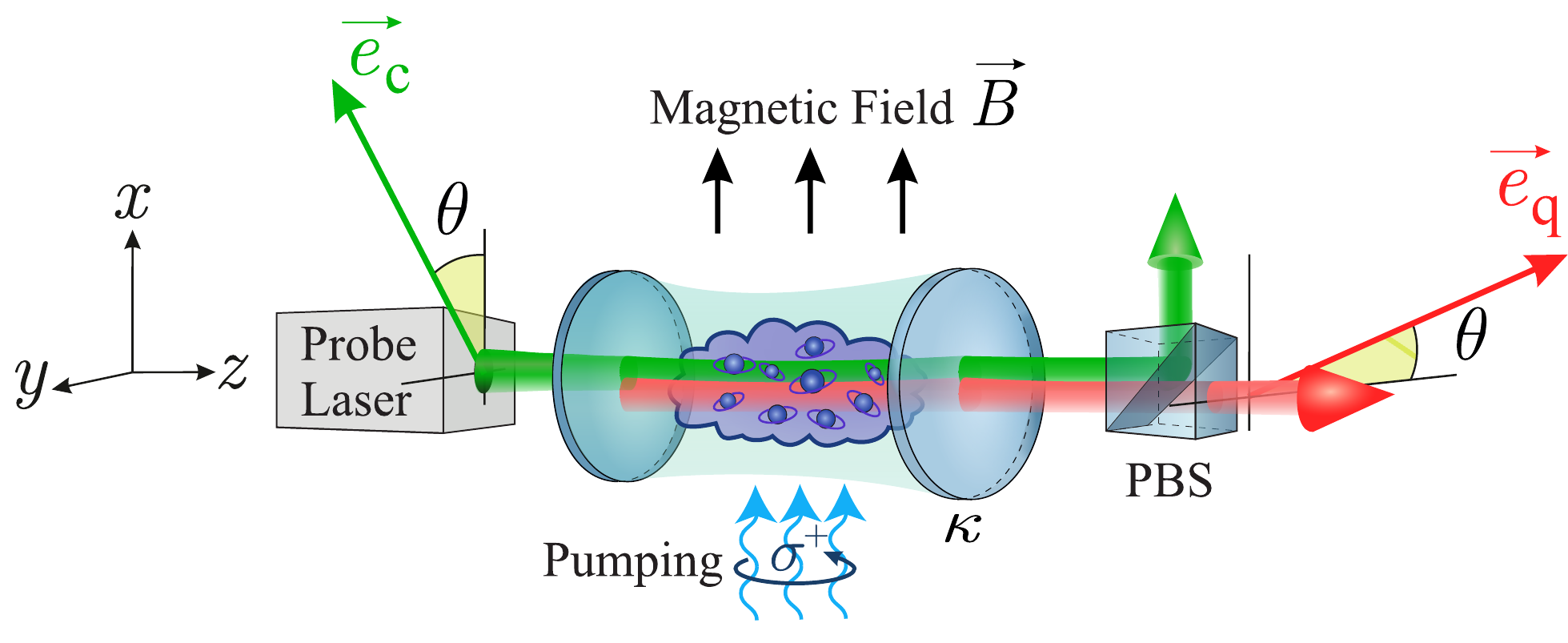}
\caption{Ensemble of Alkali atoms subject to continuous optical pumping along the direction of a homogeneous magnetic field $\vec{B}$. An off-resonant probe laser propagates along the transverse direction with linear polarization enclosing an angle $\theta$ with the mean atomic polarization. At large optical depth, collective emission generates photons in the orthogonal light polarization.}
\label{fig:Eugenes setup}
\end{figure}

Here, we consider the setup shown in figure~\ref{fig:Eugenes setup}. An ensemble of Alkali atoms is subject to optical pumping and to continuous, off-resonant probing of spin polarization transverse to the direction of mean polarization.  The treatment will closely follow that in \cite{hammerer_quantum_2010,hammerer_quantum_2006}, but extend it in two aspects: first, instead of pulsed, continuous pump and probe fields will be treated and second, the possibility of collective emissions instead of scattering of independent atoms will be considered.
We first develop the corresponding master equation in section~\ref{sec:drivmastereq} and then apply it to the examples of atoms with ground state $F=1$ in section~\ref{sec:Three level atom} and $F=4$ in section~\ref{sec:Cs results}. 
These applications will demonstrate the close connection to the model of the generalised superradiant laser introduced in the previous section.

\subsection{Master equation of continuously pumped and probed atomic ensembles}\label{sec:drivmastereq}

We consider $N$ Alkali atoms which are  continuously probed by  an off-resonant  laser of wavelength $\lambda_{\text{c}}$ propagating in $z$ direction with linear polarization enclosing an angle  $\theta$  relative to the $x$-axis, the axis of mean atomic polarization, cf. figure~\ref{fig:Eugenes setup}. The laser couples off-resonantly to one of the atomic $D$-lines with ground state spin $F$ and excited state spins $F'$. 
The respective Zeeman states will be denoted by $\ket{F,m_F}$ and $\ket{F',m_{F'}}$. We assume a spatial distribution of atoms exhibiting a large optical depth $D=\frac{N\sigma_0}{A}$ along the axis of the probe field. Here, $\sigma_0=\frac{3\lambda^2}{2\pi}$ is the scattering cross section on resonance, and $A$ is the beam cross section. In this limit, the scattering of photons in the $z$ direction occurs in the same spatial mode, which we model here by a cavity mode with linewidth $\kappa$ to which all atoms couple equally \cite{loudon_quantum_2000}. This (virtual) cavity mode is then adiabatically elimated in the limit $\kappa\rightarrow\infty$, which yields a master equation for the atoms that represents collective emissions in the $z$ direction in free space. Scattering in all other directions is non-collective, and will be covered by suitable Lindblad terms in the master equation. Regarding motion of atom, we will follow the approximations of \cite{hammerer_quantum_2010} suitable for treating an ensemble of thermal atoms in a cell. Through thermal averaging, the motion of the atoms is almost decoupled from their spin and the forward-scattered photons.

Our starting point is the master equation for $N$ atoms interacting with the electromagnetic field in dipole approximation. In the electromagnetic field we distinguish the forward scattering modes, which are modelled as a running wave cavity, and all other field 
modes,
\begin{align}
\dot \rho &= \frac{1}{\i\hbar}\C{H_\at+H_\cav+H_{\field} + H_{\Int}}\rho   +     \kappaa \L{ \a}.   
\label{eq:ME 3d, scatter, upper levels}
\end{align}
The Hamiltonians are
\begin{align}
H_\at &= \hbar \sum_{i=1}^N  \Bbreak{\sum_{F',m_{F'}}    \omega'_{F',m_{F'}} \proj[i]{F',m_{F'}} 
\breaksign+\sumset[m_F]    \omega_{F,m_F}\proj[i]{F,m_{F}} }
\notag\\
H_\cav &=     \hbar  \omegaL    \ad \a
\notag\\
H_\field &= \hbar  \sum_\lambda   \int \intd{\bm k}  \omega_{\bm k}    \ad_{\bm k,\lambda} \a_{\bm k,\lambda}    
\notag\\
H_\Int &=   \sum_{i=1}^N \sum_{F'}\E^{-}(\bm{r}_i,t)\bm d^-_{i,FF'} +\hc
\notag
.
\end{align}
We expand the electric field into its coherent ($\mathds{C}$-number) component, the (quantized) cavity field orthogonally polarized to it, and all other field modes, which are, respectively,
\begin{align}
    \E^{-}(\bm{r},t)&=\Eclassical^-(z,t)  +  \E_\cav^- (z) + \bm E_\field^-(\bm{r}),\\
    \Eclassical^-(z,t) &= \rho_{\text{c}}  \sqrt{ \Phi}  \e{-\i \B{ k_{\text{c}} z-\omegaL t  } } \bm e_\text{c},\\
    \E_{\cav}^-(z,t) &=  \rho_{q}  \ad  \e{-\i k_{\text{c}} z }      \bm e_\text{q},\\
    \bm E_{\mathrm{field}}^-(\bm{r}) &= \sum_\lambda \int \intd{\bm k} \rho_\omega  \ad_{\bm k,\lambda} \e{-\i \bm k\bm{r}} \bm e_{\bm k,\lambda}.
\label{eq:E_se initial}
\end{align}
Here, $\rho_{\text{c}} = \sqrt{\frac{\hbar \omegac}{2 \epsilon_0 c A}}$, $\rho_{\text{q}}=\sqrt{ {\kappaa  }{}}  \fracsmall{\rho_{\text{c}} }{2}$, and $\rho_\omega = \sqrt{\frac{\hbar \omega}{2 \epsilon_0 (2\pi)^3}}$ is the electrical field per photon for classical, cavity and free field, respectively. $\omegac$ is the laser frequency, $k_{\text{c}}$ its wave number, $\Phi$ the  photonflux, 
and $\bm e_\text{c} = \irow{ \cos \theta & \sin \theta & 0 }\vect$ the linear polarization of the laser field. Regarding the forward scattered quantum field (\textit{i.e.} the cavity field), $\bm e_\text{q} = \irow{ -\sin \theta & \cos \theta & 0 }\vect$ is the linear polarization vector orthogonal to $\bm e_\text{c}$, and $\kappa$ the cavity line width. Since we are eventually interested in the free space limit $\kappa\rightarrow\infty$, we take the cavity resonance frequency $\omegaL$ to be identical to the laser frequency. The dipole operator in $H_\mathrm{int}$ is expanded as $  \bm d_{i} =\bm d^+_{i,F'F}  +  \bm d^-_{i,F'F} $ as $\bm d^+_{i,F'F} =  \pi_i^{F'} \bm d_i   \pi_i^{F}$, where $\pi_i^F=\sum_{m_F}\proj[i]{F,m_F}$ are projectors in the spin-$F$-subspace of atom $i$, and $\bm d^-_{i,F'F}  = (\bm d^+_{i,F'F} )\dg$.

In a first step, we consider the dispersive limit of light-matter interaction where the detuning $\Delta$ of the laser frequency $\omegac$ from the closest atomic $F\leftrightarrow F'$ transition is large, and only resonant two photon transitions can occur. In this limit, the excited states can be adiabatically eliminated \cite{reiter_effective_2012}. In the same step, we eliminate the field modes in Born-Markov approximation \cite{breuer_theory_2007}. This results in a master equation for the ground state spins $F$ and the cavity mode, covering forward scattering of photons,
\begin{align}
\dot \rho &= \frac{1}{\i\hbar}\C{ H_{\at,g} +H_\cav +H^{\text{eff}}_\mathrm{int}}\rho
 + \kappaa \L{ \a}  \nonumber\\
 & \quad
 +\sum_{i=1}^N \sum_{\mu=1}^3 \L{L_{\at,i,\mu}^{\text{eff}}}.
 \label{eq:reiter formula} 
\end{align}
From the atomic  Hamiltonian $H_{\at}$ only the   ground state manifold remains,
\begin{align*}
H_{\at,g} &:=   \hbar \sum_{i=1}^N  \sumset[m_F]    \omega_{F,m_F}   \proj[i]{F,m_F}      
.
\end{align*}
Here the effective interaction Hamiltonian for the ground state spins with light is~\cite{reiter_effective_2012}
\begin{align*} 
H^{\text{eff}}_{\mathrm{int}} & 
\approx
    -    \sum_{i=1}^N \B{        \B{ \E_{\mathrm{cav}}^- (z_i,t)}\vect     \calT     \Ebmclassical^+(z_i,t)    +\hc}   \label{eq:Hinteff}
,
\end{align*}
where we use the polarizability tensor
\begin{align}
\calT  &:= \frac{  \abs{\dmaxlevels}^2 }{  \hbar \Delta   }   \sumset[k=0]^2  {s_k}  \That{k}
\end{align}
with scalar, vector and tensor polarizability operators \cite{geremia_tensor_2006,brink_angular_1994}
\begin{align*}
\That{0}  &=     -\frac{1}{\sqrt 3}  \1_i,
\\
\That{1} &=  \frac{\i}{\sqrt 2} \Fbm \bm\times,    
\\
\That{2} &=   \frac{1}{2} \B{2 \Fbm \tensor \Fbm + \i \Fbm \bm\times . - \frac 2 3   \B{\Fbm}^2 \1_i } .
\end{align*}
$d$ denotes the reduced dipole matrix element (in the convention of \cite{brink_angular_1994}), and $s_k$ are dimensionless, real coefficients which depend on the detuning (see (4.43) in \cite{roth_collective_2018}). For detunings much larger than the excited states' hyperfine splitting, the tensor polarizability does not contribute, $s_2\rightarrow 0$ \cite{julsgaard_brian_2003_}. In the effective light matter interaction we keep terms linear in the coherent field, and drop terms which are quadratic in the coherent field (Stark shift of atomic levels) or in the quantum field (no mean field enhancement). The Stark shift is dropped here for simplicity, but could be easily taken into account in this framework. We note that in the Hamiltonian in Eq.~\eqref{eq:Hinteff} the atomic coordinates drop out, such that the atomic positions decouple from the dynamics. This is due to the fact that we consider forward scattering only.

The Lindblad terms in the second line of Eq.~\eqref{eq:reiter formula} account for individual spontaneous emission of each atom. The jump operators can be conveniently labelled in a Cartesian basis with index $\mu=1,2,3$,
\begin{align*}  
L_{\mathrm{at},i,\mu}^{\text{eff}} &
\approx \sqrt{\gamma'}   \B{\calT   \Ebmclassical^+       }_\mu
,
\end{align*}
where we define $  \gamma'=  \frac{  \omegaL^3}{6 \pi \hbar \epsilon_0 c^3}$. We note that, due to the structure of Lindblad terms, the second line of Eq.~\eqref{eq:reiter formula} is actually basis independent. A convenient choice will be to use $\Curly{\bm{e}_\text{c},\bm{e}_\text{q},\bm{e}_\text{z}}$.

In the next step we eliminate the cavity field in the free space limit based on the methods of \cite{breuer_theory_2007,gardiner_quantum_2004}. The resulting master equation, written in a rotating frame with respect to $H_{\at,g}$, in the limit $\kappa\rightarrow\infty$ is
\begin{align}
\dot \rho &=    \sum_{\omega}  \Bbreak{   \gamma_{\mathrm{dec}}   \sum_{i=1}^N  \sumset[\mu=c,q,z]    \L{\V{\mu}(\omega)} 
 + \gammathree\,  \L{    \V[]{q}(\omega) }    } \nonumber\\
\breaksign+ w \sum_{i=1}^N \L{\F^+}
\label{eq:Final Master equation 2}
\end{align} 
We introduce here the dimensionless jump operators
\begin{align}
V^{\text{q}}(\omega)&=\sum_{i=1}^N V^{\text{q}}_{i}(\omega),
\label{eq:dimensionless jump operators Vik}\\
\V{\mu}(\omega) &= \sumset[m_F,m_F'\\ \omega_{m_F}-  \omega_{m_F'}  = \omega] \proj{m_F} \V{\mu}  \proj{m_F'}
\label{eq:eigenoperator decomposition}
\end{align}
for atom $i$ for $\mu\in\Curly{\text{c},\text{q},\text{z}}$.
The sum is over all pairs $(m_F,m'_{F})$ with a given energy splitting $\hbar\omega=\hbar(\omega_{m_F}-\omega_{m'_F})$, and
\begin{align} 
\V{\mu} &:=         \sumset[k=0]^2  {s_k}  \bm e_{\mu}\vect \That{k}  \bm e_\text{c}.
\label{eq:V_i_mu}
\end{align}
In Eq.~\eqref{eq:Final Master equation 2} we introduced the \textit{decoherence rate due to spontaneous emission},
\begin{align*}
\gamma_{\mathrm{dec}}  = \Phi\gamma'  \B{ \frac{  \abs{\dmaxlevels}^2 } {  \hbar \Delta   }  \rho_{\text{c}}    }^2=\frac{\Phi}{8}\frac{ \sigma_0}{A}\B{\frac{\gamma_0}{\Delta}}^2,
\end{align*}
and the \textit{rate of collective forward scattering},
\begin{align}
\gammathree &=   \Phi \B{\frac{  \omegac   \abs{\dmaxlevels}^2  }{2 \epsilon_0 c A  \hbar \Delta  }     }^2  
=\frac{\Phi}{16}\B{\frac{ \sigma_0}{A}}^2\B{\frac{\gamma_0}{\Delta}}^2
\label{eq:Gamma''234 definitions in bad cavity limit}
.
\end{align}
We use here the spontaneous emission rate $\gamma_0=\frac{\omegac^3|d|^2}{3\pi\epsilon_0\hbar c^3}$. We note that due to the collective nature of the jump term associated with collective scattering, the effective rate of these processes is $N\gamma$. Therefore, the relative strength of collective scattering with respect to decoherence due to spontaneous emission, $\frac{N\gamma}{\gamma_\mathrm{dec}}= \frac{D}{2}$, becomes large  for sufficiently large optical depth.
 
Furthermore, we add in the last line of Eq.~\eqref{eq:Final Master equation 2} a Lindblad term  accounting for optical pumping to the ground state with $m_F=F$. As explained earlier, we employ a phenomenological description for this process, as our main aim here is to provide a microscopic picture for the non-collective and collective effects of the continuous probe. The microscopic theory of optical pumping is of course well established, and can in principle be used to give a more realistic account than the minimal model used here. The master equation~\eqref{eq:Final Master equation 2} is the main result of this section. For more details on its derivation we refer to~\cite{roth_collective_2018}.

It is instructive to consider in more detail the form of the  jump operator  in \eqref{eq:V_i_mu}
\begin{align}
V^{\text{q}}_i &
=  { \i \frac{s_1}{ 2}  \B{\F^--\F^+}
   -  s_2 \B{
  \frac{\i \cos(2\theta)   }{\sqrt 2}  W_1 	
 + \frac{\sin(2\theta) }{4} W_2	 }   }
 \label{eq:full jump op}
\end{align}
 occurring in the collective jump term in \eqref{eq:Final Master equation 2},  where we defined the operators
\begin{align*}
W_1 &:=   {  \B{ \F^0  +  \frac{1}{2}}\F^-+ \B{  \F^0 -  \frac{1}{2} }\F^+ }
,
\\
W_2 &:=  {3 \B{\F^0}^2   -\B{\Fbm}^2 +  \B{\F^-}^2 + \B{\F^+}^2}
.
\end{align*}
The operator $W_1$ collects processes which change $m$ by $\pm 1$, and $W_2$ contains changes by $0$ or $\pm 2$. We emphasize that the $\theta$-dependence is an effect of the tensor component $\That[]{2}$ in the  polarizability tensor.


\definecolor{pumpcolor}{rgb}{0,.7,0}
\definecolor{bscolor}{rgb}{.6,.6,.6}
\definecolor{lasercolor}{rgb}{1,0,0}
\definecolor{quantumcolor}{rgb}{0,0,1}


\subsection{Ground-state  spin $F=1$}
\label{sec:Three level atom}

\begin{figure}
\centering
\includegraphics[width=0.9\linewidth]{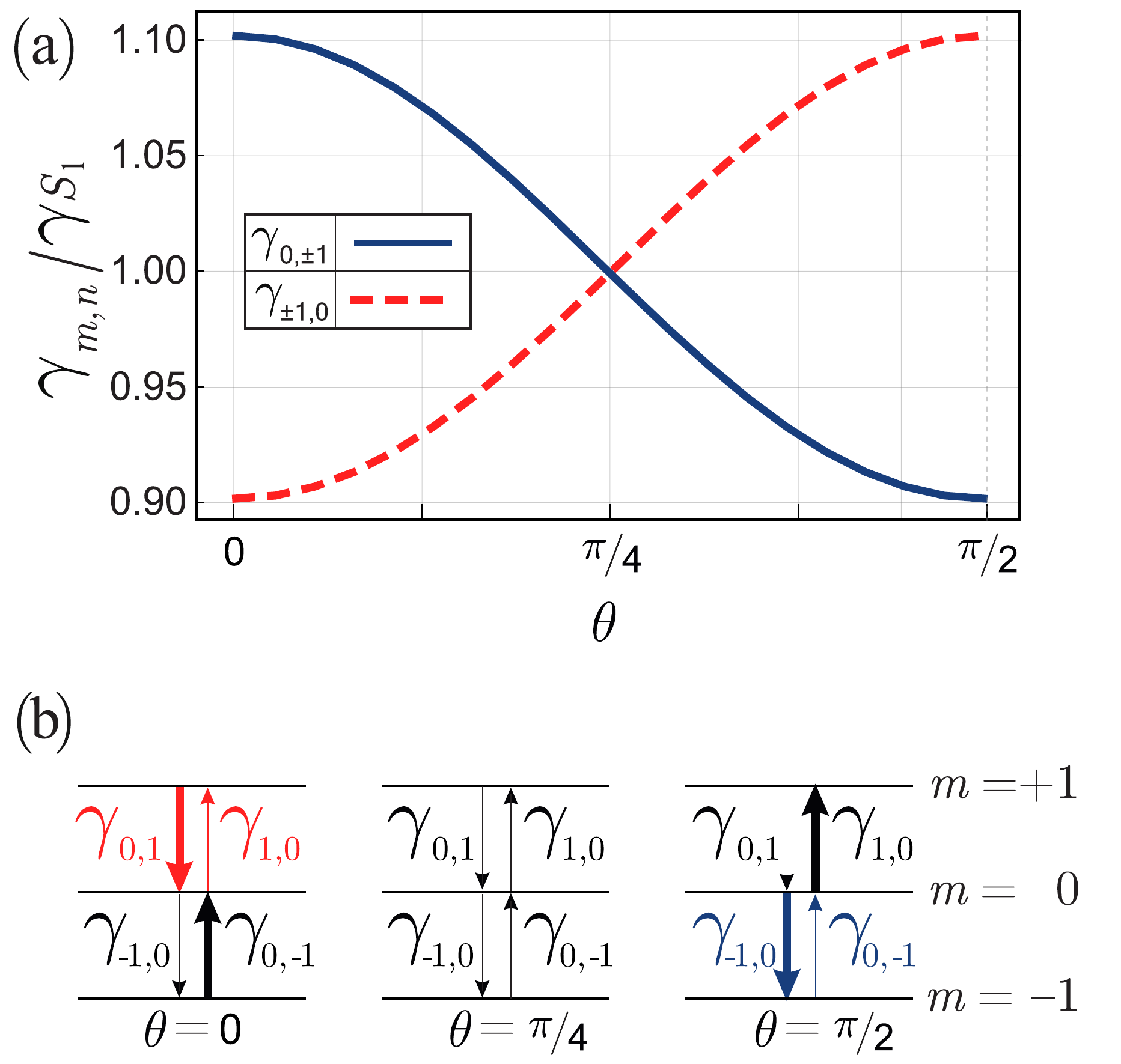} 
\caption{(a) Transition rates $\gpm{m}{n}$ versus angle $\theta$, for a relative weight of tensor to vector polarizability $\epsilon=0.1$. Rates of transitions in opposite direction and involving different levels are identical, i.e., $\gpm{0}{1}=\gpm{0}{-1}$ and $\gpm{-1}{0}=\gpm{1}{0}$. (b) Transition rates $\gpm{m}{n}$ between the different ground state levels $m$ for angles $\theta=0,\pi/4, \pi/2$. The thickness of the line represents a measure for the transition strength. Transition fulfilling the conditions for superradiant lasing are shown in red and blue.}
\label{fig:three level rates plot}
\end{figure}

We will now evaluate the master equation in Eq.~\eqref{eq:Final Master equation 2} for the case of spin $F=1$. In order to highlight the most important features more clearly, we deliberately omit the $W_2$ components in the jump operators in Eq.~\eqref{eq:full jump op} for now. With this simplification, the master equation becomes
\begin{align}
\dot \rho &=\frac{1}{i}  \sumset[i=1]^N  \C{ \sumset[m=-1]^1  \omega_m \proj[i]{m} }{\rho}
\breaksign +  \gamma    \L{V^+(\theta)}    + \gamma    \L{V^-(\theta)}
\breaksign +   w_+  \sumset[i=1]^N \L{\F^+} +   w_-  \sumset[i=1]^N \L{\F^-}.
\label{eq:three level me}
\end{align}
Here, the first term on the right hand side accounts for the splitting of the levels $\ket{m}$ in the external magnetic field with Zeemann energies $\omega_m$ where now $m=-1,0,1$.
The terms in the second line represent the effect of collective scattering of photons in the $z$-direction. The \cj operators depend on the angle $\theta$ between the polarizations of atoms and light, and are given by \begin{align}\label{eq:three levels def V^+- collective}
V^\pm(\theta) = \sumset[i=1]^N V^\pm_i(\theta)
\end{align}
with single atom operators
\begin{align}
V^\pm_i(\theta) &=  s_1\B{1 + \epsilon \cos(2\theta) \B{\mp \F^0 + \1/2}   } \F^\pm.
\label{eq:three levels def V^+-}
\end{align}
We define $\epsilon = \sqrt 2  \abs{ \fracsmall{s_2 }{ s_1}}$ measuring the relative weight of the ground states' tensor to vector polarizability. In the limit of large detuning $\epsilon$ vanishes asymptotically. The terms in the last line describe individual optical pumping and depumping at rate $w_\pm$, respectively. As in the case of the generalized superradiant laser in section \ref{sec:Generalized Superradiant Laser}, we restrict the analysis to $w_+>w_-$. The collective jump operators are associated with transitions between Zeeman states $\ket{n}$ to $\ket{m}$ where $\Delta m = m-n=\pm 1$ for $V^\pm(\theta)$, respectively. It will be useful to define the single-atom transition rates for these transitions
\begin{align}
\gpm{m}{n} &= \gamma \abs{\bra{m} V^{m-n}_i(\theta) \ket{n}}^2\label{eq:gammamn}\\
&=\gamma s_1 \B{ {1 + \epsilon \cos(2\theta) \B{\mp m + 1/2}   } }^2\nonumber
\end{align}
It can be seen that the angle $\theta$ controls the balance between $\Delta m=\pm 1$ transitions. Figure~\ref{fig:three level rates plot}(a) illustrates how the relative weight of $\gamma_{0,\pm1}$ and $\gamma_{\pm1,0}$ shifts with $\theta$. From the discussion of the generalized superradiant laser model in section~\ref{sec:Generalized Superradiant Laser}, it should be expected that the relative weight crucially determines the regimes of superradiance, as shown schematically in figure~\ref{fig:three level rates plot}(b).

\begin{figure}[t]\centering
\includegraphics[height=5.0cm]{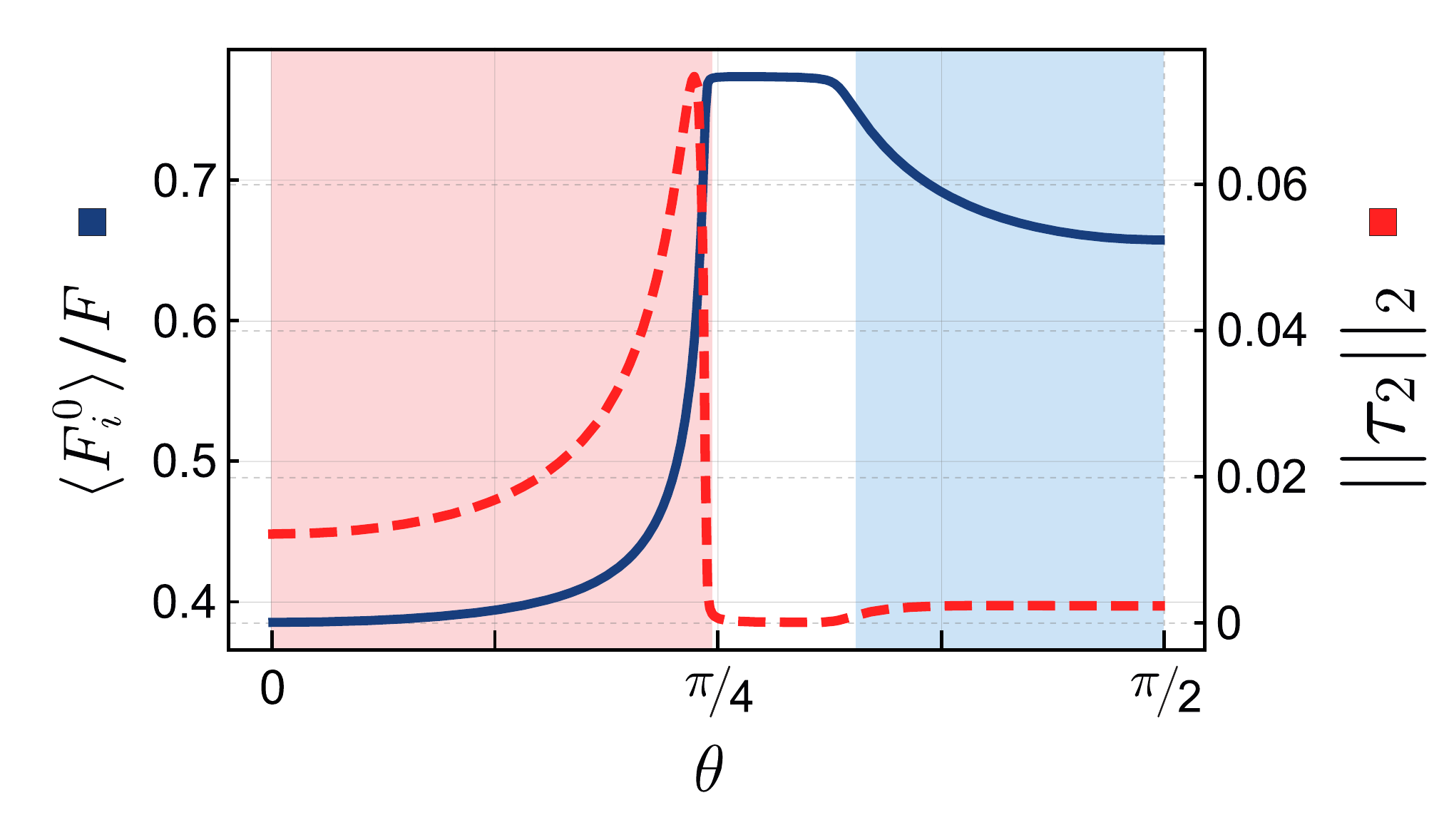}
\caption{Polarization  $\Meansmall{\F^0}/F$, and norm of two-atom correlations  $\normsmall[2]{\tau_2}$  over  the angle $\theta$, where $ \tau_2 :=\rho_2-\rho_1\otimes\rho_1$ and $\rho_n$ is the reduced density matrix of $n$ atoms. Red and blue shaded regions correspond to  $\normsmall[2]{\tau_2}>10^{-3}$  indicating significant two-atom correlations and associated lasing.  The parameters are chosen the following way: For a fixed $N\gamma$ and $N=2\cdot 10^5$ we need a small \sdj rate $ w_-=N\gamma/1000$ to be in a regime of significant collective effects. The \suj rate follows as  $w_+=5 w_-$ to create a  significant population inversion.
}
\label{fig:three level Polarization}
\end{figure}

\begin{figure}[t]\centering
\includegraphics[height=5.3cm]{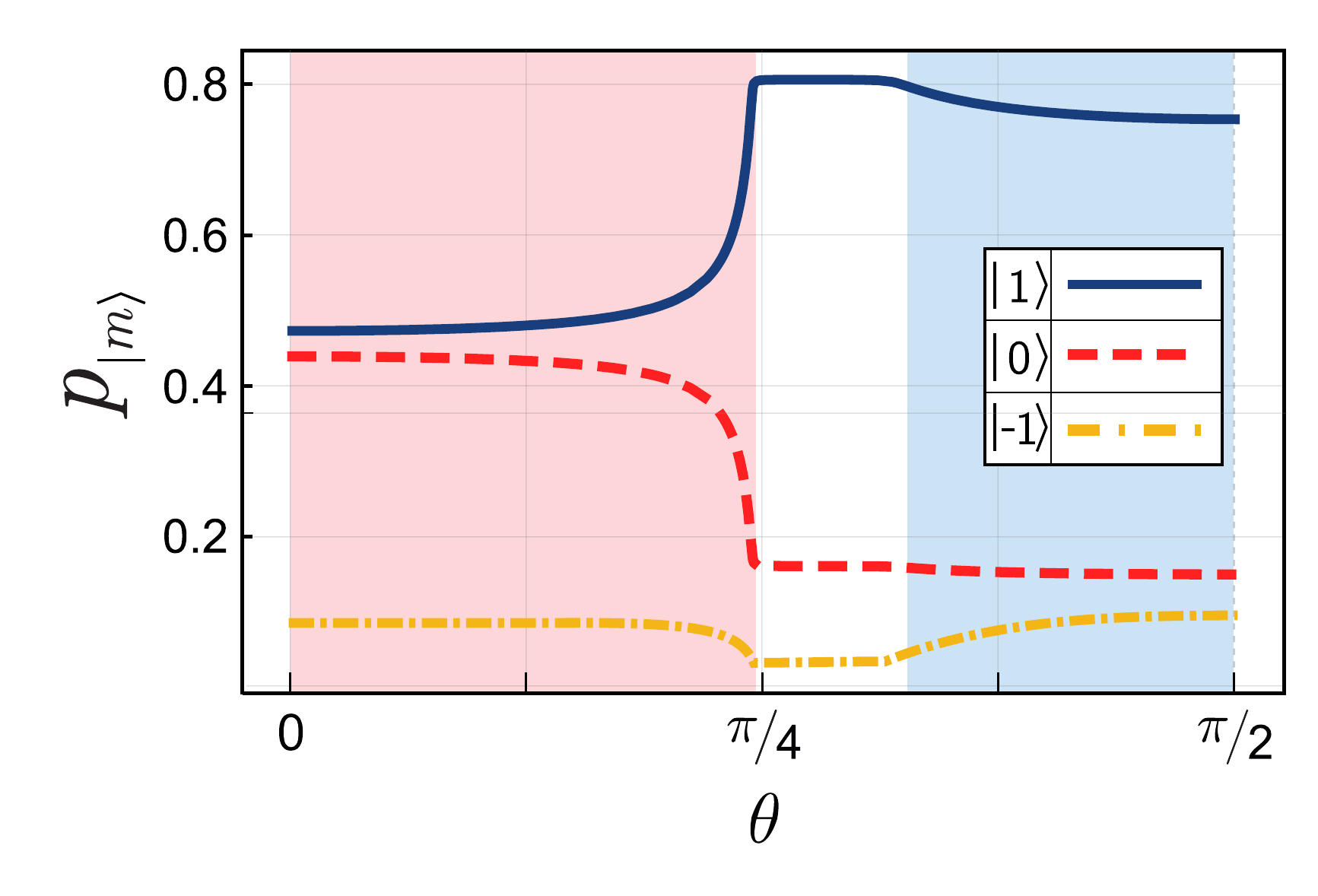}
\caption{Population $p_{|m\rangle}$ of levels $\ket{m}$ versus $\theta$. Populations experience significant redistribution in the lasing regimes (blue and red shaded areas) as compared to non-lasing regime (white area) where single atom physics prevails. The parameters are the same as in figure~\ref{fig:three level Polarization}
}
\label{fig:three level Population}
\end{figure}

As in the previous sections, the master equation~\eqref{eq:three level me} is solved for the steady state in a cumulant expansion. For this purpose, the master equation is expanded in an operator basis (with elements $A^\alpha_i$ for particle $i$), and $3$-particle correlators are approximated as $ \Mean[]{A_1^{\alpha_1}A_2^{\alpha_2}A_3^{\alpha_3}}  \approx \Mean[]{A_1^{\alpha_1}A_2^{\alpha_2}}\Mean[]{A_3^{\alpha_3}}
+\Mean[]{A_1^{\alpha_1}A_3^{\alpha_3}}\Mean[]{A_2^{\alpha_2}} + \Mean[]{A_2^{\alpha_2}A_3^{\alpha_3}}\Mean[]{A_1^{\alpha_1}}
-2\Mean[]{A_1^{\alpha_1}}\Mean[]{A_2^{\alpha_2}}\Mean[]{A_3^{\alpha_3}} $. From this approximate solution we can extract information on single particle observables such as level populations and mean polarization, as well as on the magnitude of two-particle correlations. The latter we quantify by the norm $\normsmall[2]{\tau_2}$ of $ \tau_2 =\rho_2-\rho_1\otimes\rho_1$, where $\rho_n$ denotes the $n$-body reduced density operator. The dependence of these quantities on the angle $\theta$ are shown in figure~\ref{fig:three level Polarization} and figure~\ref{fig:three level Population}.

\begin{figure}[t]\centering
\includegraphics[width=1.1\linewidth, trim=3cm 0 0 0]{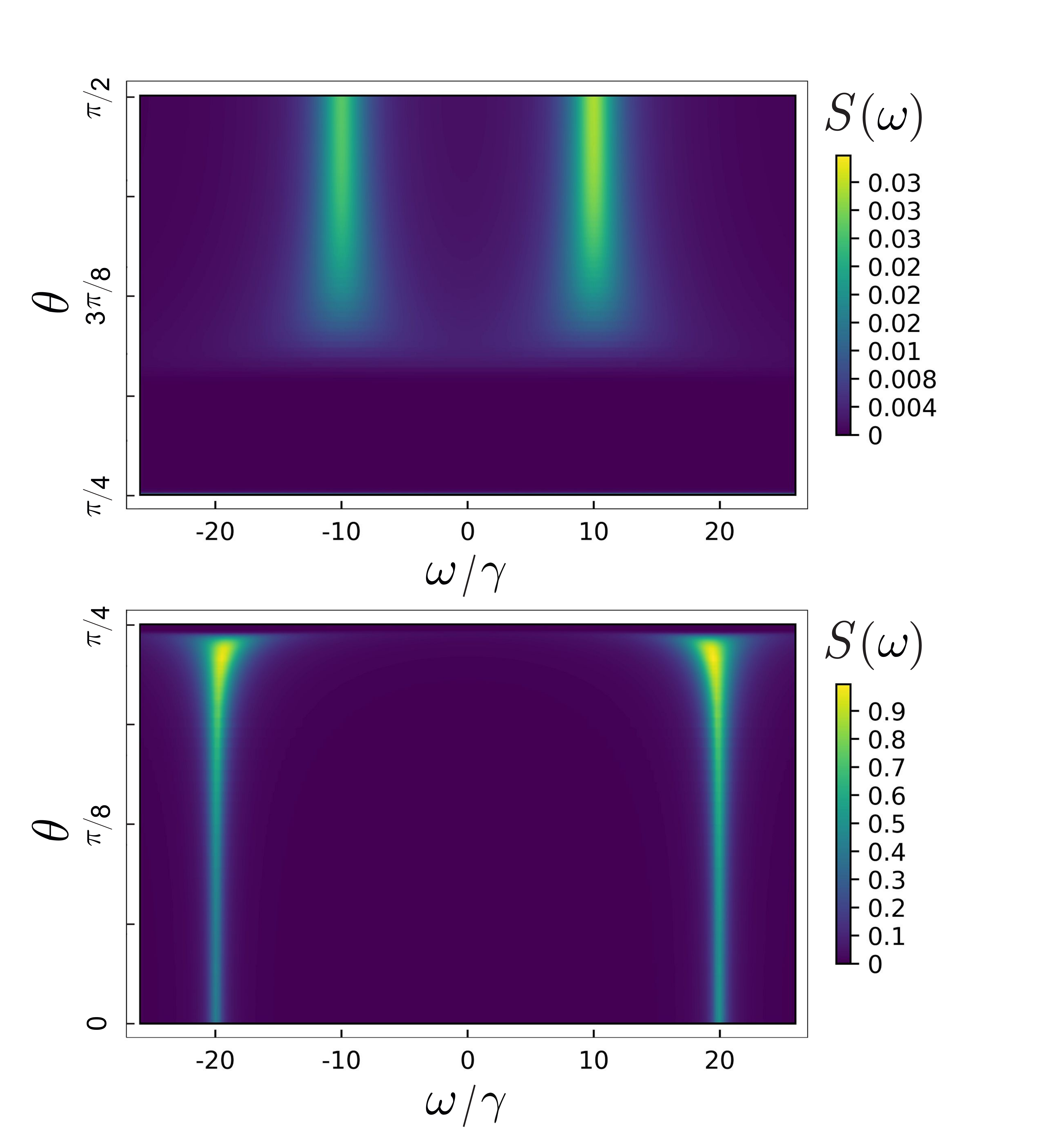}
\caption{Spectrum $S(\omega)$ of collectively scattered light versus frequency $\omega$ on the x-axis and  angle $\theta$ on the y-axis (separated in two plots due to different color scales). The lower plot  shows the two Lorentz peaks at $\omega=\pm 20\gamma $ associated with a superradiant transition on the $0\leftrightarrow 1$ levels, while the upper plot has the Lorentz peaks at $\omega=\pm 10\gamma $ associated with a superradiant transition on the $-1\leftrightarrow 0$ levels. The intensity maxima reflect the steady state populations in the respective levels. The parameters are identical to figure~\ref{fig:three level Polarization}.}
\label{fig:three level Spectrum}
\end{figure}

The mean polarization  $\Meansmall{\F^0}/F$  along the $x$-direction strongly depends on the parameter $\theta$ as a result of the interplay between the optical pumping along $x$ and quantum jumps described by the \cj operators $ V^\pm_i(\theta)$. We can understand this behavior by considering each transition in figure~\ref{fig:three level rates plot}(b) involving only two levels and comparing it with the  condition for superradiance  \eqref{eq:Generalized superradiant laser bound} of the generalized superradiant laser.
For the upper transition $ 1 \leftrightarrow  0 $  and $\theta=0$  the \cdj with rate $\gamma_{0,1}$ is dominant, due to $\gpm[-]{0}{1}/\gpm[+]{1}{0}=\B{\fracsmallB{2+\epsilon}{2-\epsilon}}^2 >1 $. This allows   superradiance, meaning correlations between atoms build up and the atoms emit collectively such that the emitted   intensity   scales with  $N^2$.
For the upper transition  and For $\theta > \pi/4$ the \cuj[s] are dominant, due to  $\gpm[-]{0}{1}/\gpm[+]{1}{0}\leq 1 $,  meaning the superradiant condition \eqref{eq:Generalized superradiant laser bound} cannot be fulfilled.
Tuning $\theta$ between $0$ and  $ \pi/4$ gives a polarization curve in figure~\ref{fig:three level Polarization} similar to figure~\ref{fig:superradiant gamma_+}. This similarity is somewhat surprising, as the change of $\theta$ in figure~
\ref{fig:three level Polarization} entails a \textit{nonlinear} change of the \textit{both} rates $\gpm[-]{0}{1} $, $\gpm[+]{1}{0} $ (see figure \ref{fig:three level rates plot}), while in figure~\ref{fig:superradiant gamma_+} only $\gamma_+ $ is linearly changed. 
The lower transition $-1\leftrightarrow 0$ can  fulfill the superradiant condition  \eqref{eq:Generalized superradiant laser bound}  only for $\theta> \pi/4$, with a maximum dominant collective down  rate $\gpm[-]{-1}{0} $ for $\theta=\pi/2$, resulting in a polarization similar to  figure~\ref{fig:superradiant gamma_+} with inverted $x$-Axis.

For both, transitions $-1\leftrightarrow 0$ and $0\leftrightarrow 1$, superradiance implies an enhanced  \cj rate proportional to $N$, necessarily decreasing the polarization  $\Meansmall{\F^0}/F$.
The superradiant transition in the red shaded region in figure~\ref{fig:three level Population}  shifts much of the population   from $\ket{1}$ to $\ket{0}$, as is shown in figure~\ref{fig:three level Population}. The small change in population of $\ket{-1}$ is a result of the \sdj[s]  with rate $w_-$ shifting the population of $\ket{0}$ downwards. 
The superradiant transition in the blue shaded region in figure~\ref{fig:three level Population} shifts the population from $\ket{0}$ to $\ket{-1}$. The change in population of  $\ket{1}$ is a result of the \sdj[s]  with rate $w_-$  shifting  the population of $\ket{1}$ downwards.

The significant $\theta$-dependent redistribution of populations away from the fully polarized state is shown in figure~\ref{fig:three level Population}. It is also clearly visible that the red shaded   regime corresponding to superradiance of  the $0\leftrightarrow 1$ transition  involves a much larger population than the blue shaded regime  corresponding to superradiance of  the $-1\leftrightarrow 0$ transition. This will be visible also in terms of the intensity of collectively scattered photons.



\begin{figure}[t]\centering
\includegraphics[height=5.3cm]{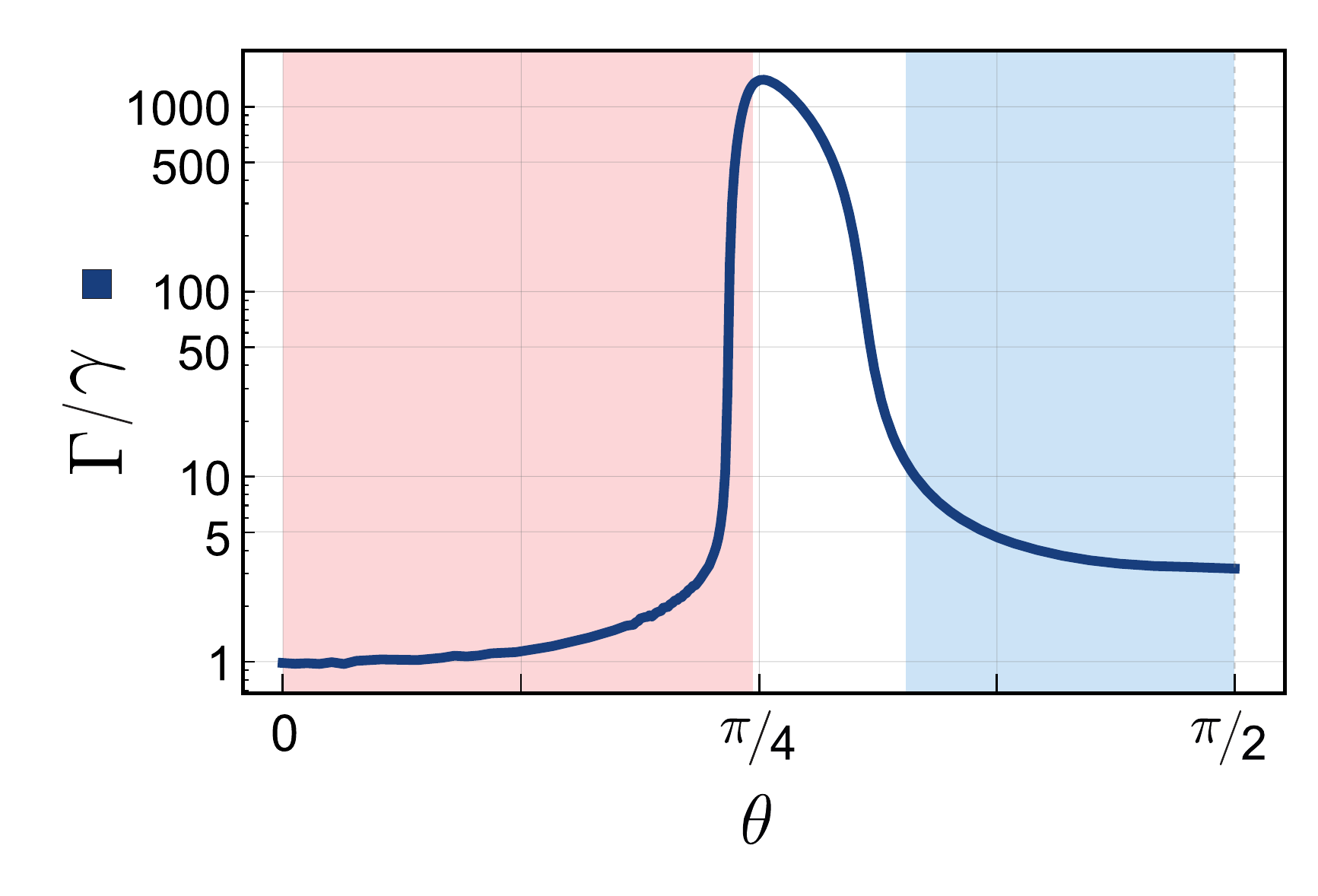}
\caption{Linewidth (full-width at half maximum) $\Gamma$  of the highest Lorentz peak versus angle $\theta$. The superradiant regimes (red and blue shaded) show a  linewidth on the order of the \cj rate $\gamma$.   The parameters are identical to figure~\ref{fig:three level Polarization}.}
\label{fig:three level gamma}
\end{figure}

The spectrum of light collectively scattered to the polarization orthogonal to the laser polarization, $S(\omega) \propto \sum_{i,j=1}^N \Ft{  \Mean{ V_i(\tau)V_j(0)}} $, follows from a Fourier transform of the atomic two-time correlation functions
\begin{align*}
&\sum_{i,j=1}^N  \Mean{ V_i(t+\tau)V_j(t)}
\=   N(N-1)\Mean{ V_2(t+\tau)V_1(t)}
   +N\Mean{ V_1(t+\tau)V_1(t)}.
 \end{align*}
Here we defined $V_i:=V_i^+ (\theta) +V_i^-(\theta) $. In order to distinguish contributions from $0\leftrightarrow 1$ and $-1\leftrightarrow 0$ transitions in the spectrum we assume a nonlinear Zeeman splitting 
with, for concreteness, $\omega_{-1}= 0$, $\omega_{0} = 10 \gamma$, and $\omega_{1} = 30 \gamma$. This particular level splitting is chosen here such that photons generated on the lower transition occur at a sideband frequency $\omega_{0}-\omega_{-1}= 10\gamma$ and for the upper transition at $ \omega_{1}-\omega_{0}= 20\gamma$.
The spectrum in figure~\ref{fig:three level Spectrum} reveals clearly that for  $0\leq \theta< \pi/4$ only the upper transition can be  superradiant and for  $\pi/4\leq \theta\leq \pi/2$ only the lower transition can be  superradiant as expected from the superradiant condition \eqref{eq:Generalized superradiant laser bound} and indicated in figure \ref{fig:three level rates plot}(b).

In addition, we extract the full-width at half maximum $\Gamma$ of the dominant Lorentz peak, as shown in figure~\ref{fig:three level gamma}. In the red and blue shaded  superradiant regions, the linewidth $\Gamma$ is on the same order of magnitude as the \cj rate $\gamma$. At $\theta = \pi/4$ the dynamics is well approximated by single atom dynamics for which the linewidth is given by $\Gamma = 2\gamma+ w_- + w_+$.

\subsection{Ground state spin $F=4$}
\label{sec:Cs results}

\begin{figure}[t]\centering
\includegraphics[height=5.3cm]{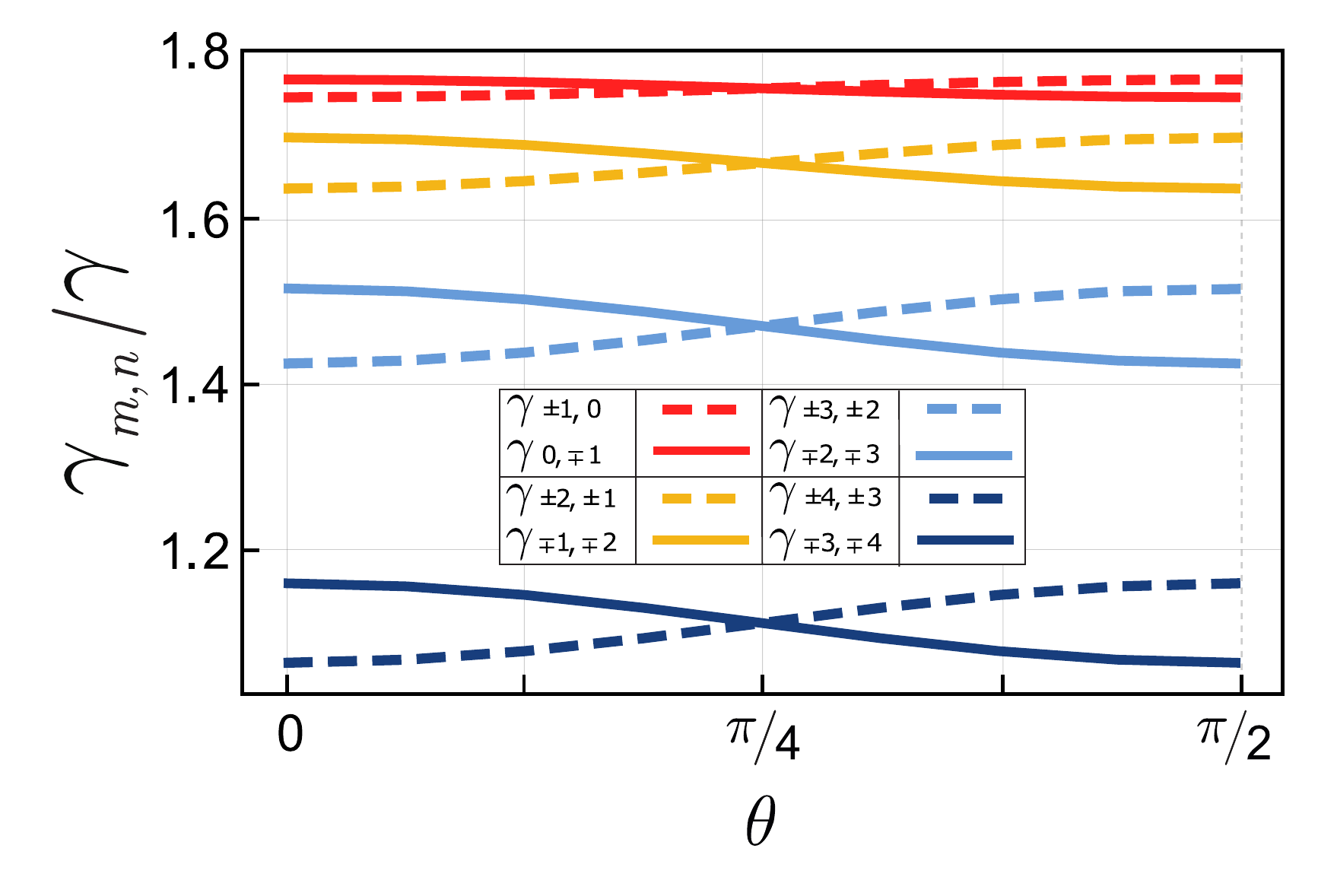}
\caption{Transition rates $\gpm{n\pm 1}{n}=\gamma \bra{n\pm 1} \V{q}\B{\pm \Omega_{\text{z}}}  \ket{n} $ (see jump operator \eqref{eq:eigenoperator decomposition}) with a Zeeman splitting $\Omega_{\text{z}} $ versus the angle $\theta$   for $F=4$ with the parameters given in figure~\ref{fig:full pol}. The rates show a similar $\theta$  dependence as in the simplified three level model in figure~\ref{fig:three level rates plot}, but their absolute value is also dependent on the hyperfine level $m$.}
\label{fig:full gammas}
\end{figure}

\begin{figure}[t]\centering
\includegraphics[height=10.6cm]{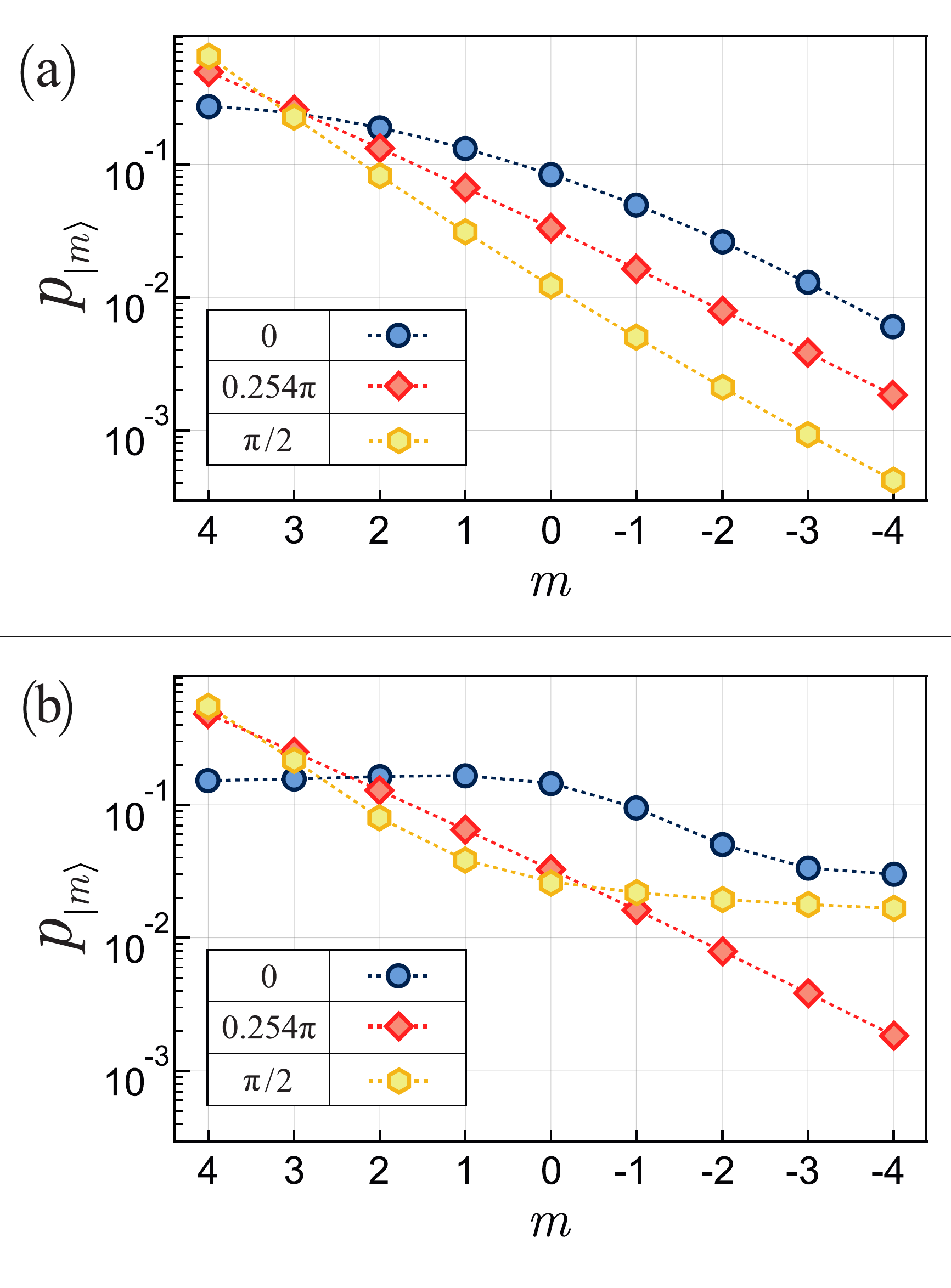}
\caption{Population $p_{|m\rangle}$ distribution for different angles  $\theta=0,0.254\pi,\pi/2$  versus the hyperfine levels  $|m\rangle$   for the same parameters as in figure~\ref{fig:full pol},   (a) has optical depth $D\approx 76$ ($N=2\cdot 10^7$ atoms), and (b)   optical depth $D\approx 1900$  ($N=5\cdot 10^8$ atoms). The lines connecting the dots are meant as a guide for the eye. 
(a) and  (b)
In the non-lasing regime  at $\theta= 0.254\pi$, the atoms are uncorrelated and exhibit an exponential distribution of populations, consistent with up and down rates independent on the level $m$.  
b) 
For  $\theta=0$ the upper transitions  become superradiant, meaning also the \cdj rate shifts the population to lower levels canceling the \suj[s] and resulting in an almost flat population distribution for $m\geq 0$; 
For  $\theta=\pi/2$ the lower transitions are superradiant  competing with the \suj[s], giving an almost flat distribution for $m\leq 0$. 
a) 
The population for $m\geq 0$ is also flattened for $\theta=0$, but not as pronounced as in (b).  For  $\theta=\pi/2$ there is no superradiance in any transition  and the population remains close to the  $\theta= 0.254\pi$ case.
}
\label{fig:full pop}
\end{figure}

\begin{figure}[t]\centering
\includegraphics[height=10.6cm]{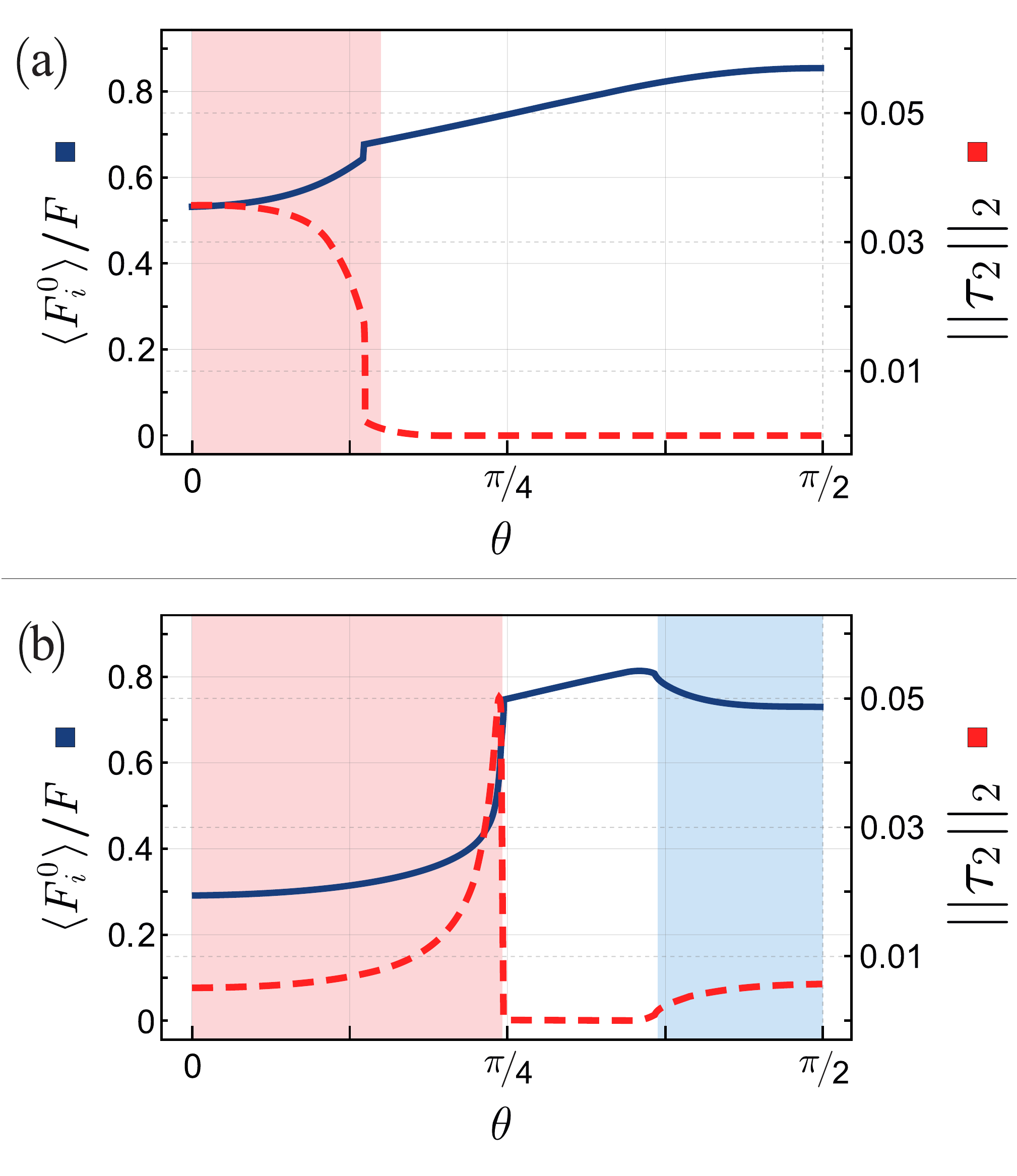}
\caption{Polarization  $\Meansmall{\F^0}/F$, and norm of   two-atom correlations   $\normsmall[2]{\tau_2}$   versus the angle $\theta$  for $F=4$, i.e., a nine-level ground state manifold. (a) has optical depth $D\approx 76$ ($N=2\cdot 10^7$ atoms), and (b)   optical depth $D\approx 1900$  ($N=5\cdot 10^8$ atoms).
The two-atom correlations  are defined as  $ \tau_2 :=\rho_2-\rho_1\otimes\rho_1$, $\rho_n$ is the reduced density matrix of $n$ atoms, and the red   and blue shaded regions correspond to   $\normsmall[2]{\tau_2}>10^{-3}$.
This figure with $F=4$ is the analog to figure \ref{fig:three level Polarization} in  the simplified  three-level system. The parameters  are the \cj rate $  \gamma/\gamma_{\text{dec}}\approx 1.9 \cdot 10^{-6}$, and  a pump rate  $ w/\gamma_{\text{dec}}\approx 5.8 \cdot 10^{-3} $. These rates correspond to a laser power $\hbar \omegaL \Phi  = 6\umW$, probe beam area $A=\B{300 \umum}^2$, laser wavelength $\lambda_L=852 \unm$, a detuning $\Delta=2\pi\cdot 3 \uGHz$, and pump rate $w=1 \ukHz$.  } 
\label{fig:full pol}
\end{figure}

Finally, we consider as an example the case of the Caesium $D_2$-line with $F=4$ and $F'=3,4,5$. 
Here, we consider the complete master equation~\eqref{eq:Final Master equation 2} without any approximation. The steady state is determined as before in cumulant expansion assuming vanishing cumulants of three or more atoms, that is, keeping only two-atom correlations.

%
 Because   the full jump operators 
 \eqref{eq:full jump op} generate    transition rates $\gpm{n\pm 1}{n}$ with similar $\theta$ dependence (see  figure  
\ref{fig:full gammas}), multiple transitions can fulfill the superradiant condition \eqref{eq:Generalized superradiant laser bound} and we   expect  multiple transitions contributing to the superradiance at the same time.
 An independent indication of which transitions are involved in the superradiance is the  population  distribution over the different hyperfine levels  plotted in figure~\ref{fig:full pop} (a) and (b). Figure \ref{fig:full pop}(b) is more instructive to see the pronounced effect of superradiance with optical depth $D\approx 1900$, while figure \ref{fig:full pop}(a) with $D\approx 76$ is more achievable. 
 For   uncorrelated atoms around $\theta\approx \pi/4$ the    \suj[s] dominate, due to the pumping rate $w$, giving an exponential population distribution in both cases.
 For $\theta=0$  figure~\ref{fig:full pop}(b) shows  the approximately flat distribution for $m\geq 0$, indicating that all upper transitions have   \cdj[s] balancing the \suj[s] dominantly created by the pumping rate $w$. This implies that all transitions between levels $m\geq 0$ are radiating collectively enhanced, i.e., are superradiant.
 For $\theta=\pi/2$  one has an inverted behavior in figure~\ref{fig:full pop}(b): The population of the upper levels is almost exponential, while the lower levels $m\leq 0$ show a flat distribution. In the lower levels the \cdj[s] are balancing the \suj[s], meaning the transitions between the levels $m\leq 0$ are radiating superradiantly.
 Figure~\ref{fig:full pop}(a) shows a less pronounced effect for $\theta=0$  and lacks any superradicance for  $\theta=\pi/2$.

Figure \ref{fig:full pol}(b) shows the same qualitative behavior for the polarization and correlations as figure \ref{fig:three level Polarization} and confirms that the choice of the   simplified jump operators  \eqref{eq:three levels def V^+-} captured the dominant effect of the full jump operators \eqref{eq:full jump op}.  The lower optical depth in figure \ref{fig:full pol}(a) compared to figure \ref{fig:full pol}(b) leads to a reduction of the   red-shaded $\theta$-range with significant  two-atom correlations  and prevents any   significant two-atom correlations for $\theta> \fracsmall{\pi}{4}$, i.e.  no superradiance on any lower transitions.

\section{Conclusion}
\label{Sec:Conclusion}
In this article we have used the methods and insights of the superradiant laser \cite{meiser_prospects_2009,kolobov_role_1993,bohnet_steady-state_2012}, specifically the  self-consistent approximation of the exact dynamics via the cumulant expansion, and applied it to the continuously pumped and off-resonantly probed atomic ensembles as present in experiments such as   \cite{krauter_entanglement_2011,moller_quantum_2017}.   
In all discussed continuously pumped and probed systems of the article we have seen parameter regimes with steady-states with significant atom-atom correlations strongly influencing observable quantities such as  the polarization.  This shows that an approximation around the single-atom steady-state, like a simple single-atom mean-field and subsequent Holstein-Primakoff transformation, would have been insufficient to capture these effects.

We see that a self-consistent approximation of the exact dynamics via the cumulant expansion   is a suitable way to derive the moment system for spin-$1/2$ atoms (see section \ref{sec:Introduction to the superradiant laser} and section \ref{sec:Generalized Superradiant Laser}) and derive analytical results, such as the superradiant lasing condition \eqref{eq:Generalized superradiant laser bound}.
For the higher spin atoms the analytical treatment becomes too tedious and one can calculate numerical results as we have shown for the spin-$1$ atoms in section \ref{sec:Three level atom} and spin-$4$ atoms in section \ref{sec:Cs results} in a setting of superradiant Raman-lasing.

The key insight in the extension of the superradiant laser in section \ref{sec:Introduction to the superradiant laser} to the generalized superradiant laser in section section \ref{sec:Generalized Superradiant Laser} shows    strong polarization dependence on the ratio $\gamma_+/\gamma_-$ of the \cuj rate $\gamma_+$ and \cdj rate $\gamma_-$ (see figure \ref{fig:superradiant gamma_+}). This  behavior then can be found again in the $F=1$ in figure \ref{fig:three level Polarization} and $F=4$ in figure \ref{fig:full pol}. Here the x-axis is the linear polarization angle $\theta$ of the probe laser, which leads to a change in the the \cuj and \cdj between neighboring excited states (see figure \ref{fig:three level rates plot} and figure \ref{fig:full gammas}) and has therefore an analog effect on the polarization.
This dramatic effect in polarization in continuously pumped and probed atomic ensembles caused by superradiance, meaning collective radiance and resulting atom-atom correlation build-up, should, in principle, be observable in experiments.

\section*{Acknowledgements}

We thank Eugene Polzik and Philipp Treutlein for discussions. KH acknowledges support from Deutsche
Forschungsgemeinschaft (DFG, German Research Foundation) under Germany’s Excellence Strategy – EXC-2123 QuantumFrontiers – 390837967, and Project-ID 274200144 – SFB 1227 (DQ-mat, project A06) through which the results in sections IIA, IIIA and IIIC were obtained. KT acknowledges financial support of the Russian Science Foundation (project no 21-72-00049) through which the results in sections IIB and IIIB were obtained.

\bibliography{bib}

\end{document}

%% file: Commands.tex
\newcommand{\B}[1]{\left(#1\right)}		
\newcommand{\Bsmall}[1]{(#1)}		
\newcommand{\Bbreak}[1]{\bracketsize{4}\{#1\bracketsize{4}\}}		
\newcommand{\Bbreakmiddle}[1]{\bracketsize{2}\{#1\bracketsize{2}\}}		
\newcommand{\C}[3][]{\left[ #2,\, #3\right]\onlyifnotempty{#1}{_{#1}}}   
\newcommand{\Cexpand}[3][]{\B{ \B{#2}\B{#3} \ifthenelse{\equal{#1}{}}{-}{#1} \B{#3}\B{#2}}  }   
\newcommand{\Curly}[1]{\left\{#1\right\}}		
\newcommand{\Eckig}[1]{\left[#1\right]}		
\newcommand{\ket}[2][]{\left|#2\right\rangle\onlyifnotempty{#1}{_{#1}}}		
\newcommand{\bra}[2][]{\left\langle#2\right|\onlyifnotempty{#1}{_{#1}}}		

\newcommand{\ketbra}[3][]{\ket{#2}\bra{#3}\onlyifnotempty{#1}{_{#1}}}		
\newcommand{\proj}[2][]{\ketbra[#1]{#2}{#2}}		
\newcommand{\Mean}[2][]{\left\langle#2\right\rangle\onlyifnotempty{#1}{_{#1}}}		
\newcommand{\Meansmall}[1]{\langle#1\rangle}		
\newcommand{\abs}[1]{\left|#1\right|}		
\newcommand{\normsmall}[2][]{||#2||\onlyifnotempty{#1}{_{#1}}}		

\newcommand{\e}[1]{e^{#1}}

\newcommand{\subsign}[2][]{#1_{\substack{#2}}}
\newcommand{\sumset}[1][]{\subsign[\sum]{#1}}

\newcommand{\fracGeneral}[4][]{\onlyifnotempty{#1}{\left.}\frac{#2 #3}{#2 #4} \onlyifnotempty{#1}{\right|_{#1}} }
\newcommand{\fracsmall}[2]{#1/#2}
\newcommand{\fracsmallB}[2]{(#1)/(#2)}

\newcommand{\D}[2][]{\fracGeneral[#1]{\mathrm{d}}{}{#2}  }

\newcommand{\irow}[1]{(\arrayy{#1})}

\newcommand{\arrayy}[1]{\begin{matrix}#1\end{matrix}}

\newcommand{\atomnu}{{\omega_{\text{eg}}}}

\renewcommand{\u}[2][]{\,#1\text{#2}}

\newcommand{\umum}{{\u[\mu]{m}}}
\newcommand{\unm}{{\u{nm}}}

\newcommand{\umW}{{\u{mW}}}

\newcommand{\ukHz}{{\u{kHz}}}

\newcommand{\uGHz}{{\u{GHz}}}

\newcommand{\myclearpage}{
\newpage
\thispagestyle{empty}
\text{}		
}

\newcommand{\insertpageifodd}{
  \ifthenelse{\isodd{\thepage}}%
    {{\newpage\text{ \thispagestyle{empty} }}}%
    {}}

\newcommand{\insertemptypageifodd}{
  \ifthenelse{\isodd{\thepage}}%
    {\myclearpage}%
    {}}

\newcommand{\define}[2][]
{\ifthenelse{\equal{#1}{}}
{\textbf{\textit{#2}}\label{def:#2}\index{#2}\xspace}
{\textbf{\textit{#1}}\label{def:#2}\index{#2@#1}\xspace}}

\newcommand{\dr}[2][]
{\ifthenelse{\equal{1}{1}}
	{\ifthenelse{\equal{#1}{}}
		{\hyperref[def:#2]{\textcolor{black}{#2}}\xspace}  
		{\hyperref[def:#2]{\textcolor{black}{#1}}\xspace}}
	{\ifthenelse{\equal{#1}{}}
		{{#2}\xspace}   
		{{#1}\xspace}}}

\newcommand{\ub}[2]{\underbrace{#1}_{\mathclap{#2}}}

\newcommand{\intd}[1]{\mathrm{d}#1 \; }

\newcommand{\dg}{^\dagger}

\renewcommand{\1}{\mathds{1}}

\renewcommand{\=}[1][=]{\notag\\ &#1}

\newcommand{\breaksign}[1]{ \notag\\  & \phantom{=}\, #1 }

\newcommand{\Ft}[2][\omega]{\mathcal F\left[#2\right]\onlyifnotempty{#1}{\B{#1}}}

\renewcommand{\L}[2][\rho]{{\mathcal D}\left[#2\right]{#1}}

\newcommand{\x}{{\hat x}}

\renewcommand{\a}{{\hat a}}

\newcommand{\ad}{\hat a^\dagger}

\newcommand{\F}[1][i]{F\onlyifnotempty{#1}{_{#1}}}
\newcommand{\Fbm}[1][i]{\bm F\onlyifnotempty{#1}{_{#1}}}
\newcommand{\Tgeneral}[3][i]{#2\onlyifnotempty{#1}{_{#1}}^{\onlyifnotempty{#3}{(#3)}}}
\newcommand{\TA}[3][]{\Tgeneral[#1]{#2}{#3}}

\newcommand{\That}[2][i]{\TA[#1]{\hat T}{#2}}
\newcommand{\calT}[1][i]{\overleftrightsmallarrow\alpha\onlyifnotempty{#1}{_{#1}}}

\newcommand{\V}[2][i]{V^{#2}\onlyifnotempty{#1}{_{#1}}}

\newcommand{\gammathree}{\gamma}

\newcommand{\onlyifnotempty}[2]{\ifthenelse{\equal{#1}{}}{}{#2}}

\renewcommand{\tensor}{\otimes\,}

\newcommand{\cj}[1][]{collective jump{#1}\xspace}
\newcommand{\cuj}[1][]{collective excitation{#1}\xspace}
\newcommand{\cdj}[1][]{collective emission{#1}\xspace}

\newcommand{\suj}[1][]{single-atom pumping{#1}\xspace}
\newcommand{\sdj}[1][]{single-atom depumping{#1}\xspace}

\newcommand{\eadr}{effective atomic decay rate\xspace}
\renewcommand{\aa}{atom-atom\xspace}

\newcommand{\vectinline}{T}
\newcommand{\vect}{^\vectinline}

\newcommand{\at}{{\mathrm{at}}}
\newcommand{\cav}{{\mathrm{cav}}}
\newcommand{\field}{{\mathrm{field}}}
\newcommand{\Int}{{\mathrm{int}}}

\newcommand{\gennu}{\nu}

\newcommand{\rme}[4][satchler]{
\ifthenelse{\equal{#1}{satchler}}
{\left\langle #2 \middle\| #3  \middle\| #4 \right\rangle  }
{
\ifthenelse{\equal{#1}{messiah}}
{\B{ #2 \middle\| #3  \middle\| #4 }}
{\text{Error! RME definition not specified}}
}
}   
\newcommand{\rmesmall}[4][satchler]{
\ifthenelse{\equal{#1}{satchler}}
{\langle #2 \| #3 \| #4 \rangle  }
{
\ifthenelse{\equal{#1}{messiah}}
{\Bsmall{ #2 \| #3  \| #4 }}
{\text{Error! RME definition not specified}}
}
}   

\renewcommand{\i}{{\text{i}}}
\newcommand{\omegaL}{\omega_\text{c}}
\newcommand{\omegac}{\omega_\text{c}}

\newcommand{\gpm}[3][\pm]{\gamma^{}_{#2,#3}}  

\newcommand{\kappaa}{\kappa} 
\newcommand{\QRT}{Quantum Regression Theorem\xspace} 

\newcommand{\dmaxlevels}{d }

\newcommand{\hc}{\text{h.c.}}

\newcommand{\cugeneral}[4][]{
\ifthenelse{\equal{#1}{}}
{#3}
{\IfSubStringInString{0}{#1}{#4}{\ub{#4}{\Eckig{#1}}}}
}

\newcommand{\bracketsize}[1]{
\ifthenelse{\equal{#1}{1}}
    {\big}
    {\ifthenelse{\equal{#1}{2}}
	{\Big}
	{\ifthenelse{\equal{#1}{3}}
	    {\bigg}
	    {\ifthenelse{\equal{#1}{4}}
		{\Bigg}
		{}	    
	    }
	}
    }
}

\newcommand{\Eclassical}{{{\bm{\mathcal{E}}}}}
\newcommand{\E}{{\bm{{E}}}}

\newcommand{\Ebmclassical}{{{\bm{\mathcal{E}}}}}

\makeatletter
\newcommand{\overleftrightsmallarrow}{\mathpalette{\overarrowsmall@\leftrightarrowfill@}}
\newcommand{\overrightsmallarrow}{\mathpalette{\overarrowsmall@\rightarrowfill@}}
\newcommand{\overleftsmallarrow}{\mathpalette{\overarrowsmall@\leftarrowfill@}}
\newcommand{\overarrowsmall@}[3]{%
  \vbox{%
    \ialign{%
      ##\crcr
      #1{\smaller@style{#2}}\crcr
      \noalign{\nointerlineskip}%
      $\m@th\hfil#2#3\hfil$\crcr
    }%
  }%
}
\def\smaller@style#1{%
  \ifx#1\displaystyle\scriptstyle\else
    \ifx#1\textstyle\scriptstyle\else
      \scriptscriptstyle
    \fi
  \fi
}
\makeatother

\newcommand{\peta}[2]{
\ifthenelse{\equal{#2}{V}}{
\eta_{#1}}{
\xi_{#1}}}